\documentclass{pasa}%
\title[Galaxy evolution studies with SPICA ]{Galaxy evolution studies with the SPace IR telescope for Cosmology and Astrophysics (SPICA):
the power of IR spectroscopy}

\author[L. Spinoglio et al.]{L. Spinoglio$^{1}$\thanks{Email: luigi.spinoglio@iaps.inaf.it}, 
A. Alonso-Herrero$^{2}$,
L. Armus$^{3}$, 
M. Baes$^{4}$, 
J. Bernard-Salas$^{5}$, 
S. Bianchi$^{6}$,  
M. Bocchio$^{6}$, 
A. Bolatto$^{7}$, 
C. Bradford$^{8}$, 
J. Braine$^{9}$,
F. J. Carrera$^{10}$, 
L. Ciesla$^{11}$,
D. L. Clements$^{12}$, 
H. Dannerbauer$^{13,14}$, 
Y. Doi$^{15}$,
A. Efstathiou$^{16}$, 
E. Egami$^{17}$, 
J. A. Fern\'andez-Ontiveros$^{1,13,14}$,
A. Ferrara$^{18}$, 
J. Fischer$^{19}$,
A. Franceschini$^{20}$, 
S. Gallerani$^{18}$, 
M. Giard$^{21}$, 
E. Gonz\'alez-Alfonso$^{22}$, 
C. Gruppioni$^{23}$, 
P. Guillard$^{24}$,
E. Hatziminaoglou$^{25}$, 
M. Imanishi$^{26}$, 
D. Ishihara$^{27}$,
N. Isobe$^{28}$, 
H. Kaneda$^{27}$,
M. Kawada$^{29}$, 
K. Kohno$^{30}$, 
J. Kwon$^{29}$, 
S. Madden$^{11}$,
M. A. Malkan$^{31}$, 
S. Marassi$^{32}$, 
H. Matsuhara$^{29}$, 
M. Matsuura$^{33}$,
G. Miniutti$^{2}$, 
K. Nagamine$^{34}$,
T. Nagao$^{35}$,
F. Najarro$^{36}$,
T. Nakagawa$^{29}$,
T. Onaka$^{37}$, 
S. Oyabu$^{27}$,
A. Pallottini$^{18,38,39}$, 
L. Piro$^{1}$,
F. Pozzi$^{40}$, 
G. Rodighiero$^{20}$,
P. Roelfsema$^{41,42}$,  
I. Sakon$^{37}$, 
P. Santini$^{32}$, 
D. Schaerer$^{43}$, 
R. Schneider$^{32,44}$, 
D. Scott$^{45}$,
S. Serjeant$^{5}$,
H. Shibai$^{35}$,                    
J.-D. T. Smith$^{46}$, 
E. Sobacchi$^{18}$, 
E. Sturm$^{47}$, 
T. Suzuki$^{27}$, 
L. Vallini$^{23,40,48}$, 
F. van der Tak$^{41,42}$, 
C. Vignali$^{40}$,
T. Yamada$^{29}$, 
T. Wada$^{29}$ \and
L. Wang$^{41,42}$\\ 
To be submitted to PASA, date: XX/XX/XX.
\normalsize
}%%%
\jid{PASA}
\doi{10.1017/pas.\the\year.xxx}
\jyear{\the\year}
\usepackage[authoryear]{natbib}
\bibpunct{(}{)}{;}{a}{}{,}
\usepackage{aas_macros}
\usepackage{hyperref} 
\hypersetup{colorlinks,citecolor=blue,linkcolor=blue,urlcolor=blue}

\begin{document}%

\begin{abstract}
IR spectroscopy in the range 12--230\,$\mu$m with the SPace IR telescope for Cosmology and Astrophysics (SPICA) will reveal the physical processes that govern the formation and evolution of galaxies and black holes through cosmic time, bridging the gap between the James Webb Space Telescope (JWST) and the new generation of Extremely Large Telescopes (ELTs) at shorter wavelengths and the Atacama Large Millimeter Array (ALMA) at longer wavelengths. 
SPICA, with its 2.5-m  telescope actively-cooled to below 8\,K, will obtain the first spectroscopic determination, in the mid-IR rest-frame, of both the star-formation rate and black hole accretion rate histories of galaxies, reaching lookback times of 12 Gyr, for large statistically significant samples. 
Densities, temperatures, radiation fields and gas-phase metallicities will be measured in dust-obscured galaxies and active galactic nuclei (AGN), sampling a large range in mass and luminosity, from faint local dwarf galaxies to luminous quasars in the distant Universe.  
AGN and starburst feedback and feeding mechanisms in distant galaxies will be uncovered through detailed measurements of molecular and atomic line profiles. SPICA's large-area deep spectrophotometric surveys will provide mid-IR spectra and continuum fluxes for unbiased samples of tens of thousands of galaxies, out to redshifts of $z$$\sim$6. Furthermore, SPICA spectroscopy will uncover the most luminous galaxies in the first few hundred million years of the Universe, through their characteristic dust and molecular hydrogen features. 
\end{abstract}
\maketitle%
{\bf Preface}

\vspace{0.5cm}
\noindent
The following set of papers describe in detail the science goals of the future Space Infrared telescope for Cosmology and Astrophysics (SPICA). The SPICA satellite will employ a 2.5-m telescope, actively cooled to around 6\,K, and a suite of mid- to far-IR spectrometers and photometric cameras, equipped with state of the art detectors. In particular the SPICA Far Infrared Instrument (SAFARI) will be a grating spectrograph with low (R=300) and medium (R$\simeq$3000--11000) resolution observing modes instantaneously covering the 35--230\,$\mu$m wavelength range. The SPICA Mid-Infrared Instrument (SMI) will have three operating modes:  a large field of view (12'$\times$10') low-resolution 17--36\,$\mu$m spectroscopic (R$\sim$50--120) and photometric camera at 34$\mu$m, a medium resolution (R$\simeq$2000)  grating spectrometer covering wavelengths of 17--36\,$\mu$m and a high-resolution echelle module (R$\simeq$28000) for the 12--18\,$\mu$m domain.  A  large field of view (80''$\times$80''), three channel, (110\,$\mu$m, 220\,$\mu$m and 350\,$\mu$m) polarimetric camera will also be part of the instrument complement. These articles will focus on some of the major scientific questions that the SPICA mission aims to address, more details about the mission and instruments can be found in  \citet{roe17}. %Roelfsema et.al. in prep.
\section{INTRODUCTION }
\label{sec:intro}

\begin{figure*}[t]
  \begin{center}
    \begin{minipage}[c]{0.67\textwidth}
% \centering
 \includegraphics[width=0.9\textwidth]{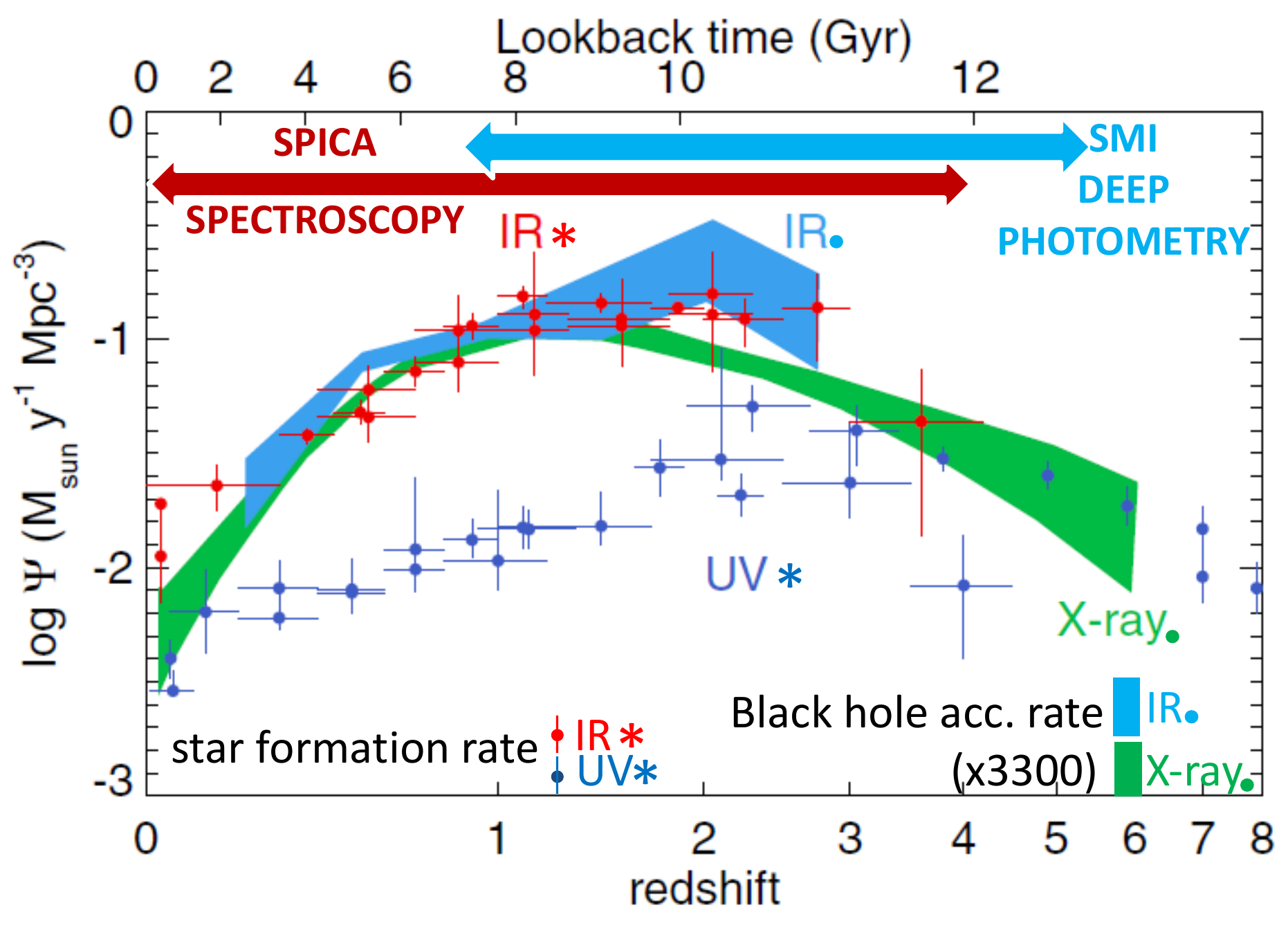}
  \end{minipage}\hfill
    \begin{minipage}[c]{0.33\textwidth}
    \caption{Estimated star-formation rate densities from the far-ultraviolet (blue points) and far-IR (red points) photometric surveys \citep[adapted from][]{2014ARA&A..52..415M}. 
%(adapted from Madau \& Dickinson, 2014). 
The estimated BHAR density, scaled up by a factor of 3300, is shown for comparison (in green shading from X-rays and light blue from the IR). 
For redshifts above $z$=2, very little IR data on the SFR density exist, making its determination rather poor.    
The redshift ranges probed with SPICA spectroscopy (up to $z$$\sim$4) and photometry (up to $z$$\sim$6) are also shown. 
    } \label{Fig1}
  \end{minipage}
%\includegraphics[width=0.8\textwidth]{Figure1.pdf}
%%\includegraphics[width=0.8\textwidth]{Figure1.pdf}
%%\includegraphics{Figure1.pdf}
%\caption{Estimated star-formation rate densities from the far-ultraviolet (FUV, blue points) and far-IR (FIR, red points) photometric surveys \citep[adapted from][]{2014ARA&A..52..415M}. 
%%(adapted from Madau \& Dickinson, 2014). 
%The estimated black hole accretion rate density, scaled up by a factor of 3300, is shown for comparison (in green shading from X-rays and light blue from the IR). 
%For redshifts above $z$=2, very little IR data on the SFR density exist, making its determination rather poor.    
%The redshift ranges probed with SPICA spectroscopy (up to $z$$\sim$4) and photometry (up to $z$$\sim$6) are also shown. 
%}\label{Fig1}
\end{center}
\end{figure*}

Over the past three decades, we have learned that at least half of the energy ever emitted by stars and accreting black holes in galaxies is absorbed by dust, and re-radiated in the infrared
\citep[e.g.][]{2001ARA&A..39..249H, 2005ARA&A..43..727L, 2008A&A...487..837F}.  %(Hauser and Dwek 2001; Lagache+2005, Franceschini et al. 2008)
We now know that the peak in the growth of galaxies %, optimally observed in IR emission, 
occurs at redshifts of $z$$\sim$1--3 \citep{1996ApJ...460L...1L, 1999MNRAS.310L...5F, 2014ARA&A..52..415M, 2016MNRAS.461.1100R}, %(Lilly et al. 1996; Franceschini et al. 1999; Madau \& Dickinson 2014; Rowan-Robinson et al. 2016), 
when the Universe was roughly 3 Gyr old --- a result achieved primarily through deep and wide-field observations with previous IR space observatories, namely the IR Astronomical Satellite  \citep[IRAS,][]{1984ApJ...278L...1N}, the IR Space Observatory  \citep[ISO,][]{1996A&A...315L..27K}, {\it Spitzer} \citep{2004ApJS..154....1W}, AKARI \citep{2007PASJ...59S.369M}, {\it Herschel} \citep{2010A&A...518L...1P} and the Wide-field IR Survey Explorer  \citep[WISE,][]{2010AJ....140.1868W}.  Despite their successes, these observatories had either small cold telescopes, or large, warm mirrors, ultimately limiting their ability to probe the physics, through spectroscopy and deep photometry, of the faintest and most distant obscured sources in our Universe. 

Due to the progress in detector performance and cryogenic cooling technologies, great advances in our ability to study the hidden, dusty Universe can be made through observations in the thermal infrared. The SPace IR telescope for Cosmology and Astrophysics \citep[SPICA,][]{2009ExA....23..193S, 2014SPIE.9143E..1IN} %Swinyard=2009, nakagawa+2014
will achieve a gain of over two orders of magnitude in spectroscopic sensitivity in the mid/far-IR compared to {\it Herschel} and {\it Spitzer}. SPICA will provide access to wavelengths well beyond those reachable with the James Webb Space Telescope  \citep[JWST,][]{2006SSRv..123..485G} and the new generation of extremely large telescopes (ELTs), and at wavelengths shortward of those accessible by the Atacama Large Millimeter/Submillimeter Array \citep[ALMA,][]{2009IEEEP..97.1463W}. It will enable the discovery and detailed study of  normal galaxies across their key phases of evolution, as well as probing the earliest forming galaxies and super-massive black holes. 

A description of the SPICA mission was originally presented in \citet{2009ExA....23..193S} and recent updates can be found in  \citet{2014SPIE.9143E..1IN, sib16}, and %nakagawa+2014 Sibthorpe et al. (2016) and 
in \citet{roe17}. 
%Roelfsema et al. (2016, in prep.). %\citet{2016EAS....75..411}
% ALL THIS DESCRIPTION HAS BEEN PUT IN THE PREFACE
%SPICA will consist of a 2.5-m primary mirror, actively cooled to below 8K.  
%There are 
The two primary instruments sharing the focal plane are the SPICA 
Far-IR Instrument (SAFARI) and the SPICA Mid-IR Instrument (SMI): 
%SAFARI includes a grating spectrometer covering  the 34--230\,$\mu$m spectral range simultaneously at a resolution {\it R}$\sim$300. The combination of the grating with a Martin-Puplett interferometer allows for observations at higher spectral resolution (1500$<${\it R}$<$11000, depending on wavelength) over the same spectral range.  
%SAFARI also includes the imaging polarimeter POL, with a field of view of 80$^{\prime \prime}$ $\times$ 80$^{\prime \prime}$ in three spectral bands centred at 110, 220 and 350\,$\mu$m.
a description of the SAFARI optical system architecture and design concept is given in \citet{2016SPIE.9904E..3UP}, while 
%SMI covers the wavelength range of 12--36\,$\mu$m, using three spectroscopic channels: low-resolution ({\it R} = 50--120, 17--36\,$\mu$m); mid-resolution ({\it R} = 1300--2300; 18--36\,$\mu$m); and high-resolution ({\it R} = 28000, 12--18\,$\mu$m). % spectroscopy. Besides the spectroscopic modes, 
%SMI can also %has an imaging capability to 
%obtain large field-of-view (10$^{\prime}\times12^{\prime}$) images at 34\,$\mu$m.   
a full description of the SMI instrument can be found in \citet{2016SPIE.9904E..2IK}. %Kaneda et al. (2016).

This article describes how %the main astrophysical questions that a mid- to far-IR space telescope, such as 
SPICA will be able to address the study of galaxy formation and evolution, while a companion article focuses on the studies of the ISM in nearby galaxies \citep{flo17}. 
It is organised as follows. Section \ref{sect2} introduces the current knowledge that has been accumulated so far in studying galaxy evolution and identifies a number of open questions that can be
addressed and answered with high sensitivity IR spectroscopic and photometric observations. We 
show in Sect. \ref{sect3} how IR spectroscopy is able to separate the two main energy production mechanisms of star formation and black hole accretion
driving galaxy evolution up to redshift of $z$=3--4.
%; in  sect.  these two mechanisms can be differentiated. 
Finally, Sect. \ref{syner} illustrates the synergies of the SPICA observations with current and future facilities at other frequencies and sect. \ref{conc} gives the conclusions. 
 
\section{The rise and fall of Galaxy Formation} \label{sect2}

The bulk of star formation and supermassive black hole (SMBH) accretion in galaxies appears to have taken place more than six billion years ago, 
dropping sharply towards the present epoch \citep[e.g.,][and references therein, see Fig. \ref{Fig1}]{2014ARA&A..52..415M}. 
The Madau \& Dickinson compilation indicates that most of the UV/optical photons, arising from hot young stars characteristic of the active star-forming areas that combine to make star-forming galaxies, have been absorbed by dust and re-readiated.  Thus as a general condition, rest-frame ultraviolet and optical observations do not access these crucial regions where gas forms massive stars and vice versa. Moreover, 
since around half of the energy emitted by stars and accreting SMBH %super-massive black holes 
is absorbed and re-emitted by dust
\citep[e.g.][]{2001ARA&A..39..249H},  %(Hauser and Dwek 2001)
understanding the physics of galaxy evolution requires IR observations of large, unbiased samples of galaxies spanning a wide range in luminosity, redshift, environment and nuclear activity. 
From {\it Spitzer} and {\it Herschel} photometric surveys the star-formation rate (SFR) and black-hole accretion rate (BHAR) density functions have been {\it estimated} through the bolometric luminosities of galaxies
\citep{2005ApJ...632..169L, 2013MNRAS.432...23G, 2014MNRAS.439.2736D}. %(le floc'h+2005, Gruppioni+13, Delveccio+2014) 
However, these estimates should be treated with caution, because they are typically based on observations of only a few, broad IR bands \citep[see, e.g.,][]{2005ApJ...630...82P}, making the IR luminosities and the relative contribution of star formation and SMBH accretion, highly uncertain.
%estimates have intrinsic limits and have to be taken with caution, %are not fully reliable, 
%because they are based on the observed integrated luminosities alone --- the contributions due to star formation and to black hole accretion have not been separated on the basis of observed physical quantities. 
This crucial separation has been attempted so far through modelling of the spectral energy distributions (SEDs) \citep[ see, e.g.,][]{2011MNRAS.414.1082M} %(Mullaney et al 2011)
and relies on model-dependent assumptions and local templates, with large uncertainties and degeneracies.  Indeed  {\it Spitzer}-  and {\it Herschel}-based studies have been successful in estimating counts and evolving galaxy luminosity functions \citep{2005ApJ...632..169L, 2005ApJ...630...82P, 2012MNRAS.424.1614O, 2013A&A...553A.132M, 2011ApJ...742...24L}, 
%(Le Floc'h et al. (2005), Pérez-González et al. (2005) Hermes---Oliver+12, PEP---Magnelli+13 and H-Atlas---Lapi+11 papers), 
but only through statistical techniques applied to bulk galaxy populations as a function of redshift.
%; {\it Herschel} was not sensitive enough to probe the physical properties of high-redshift galaxies on an object-by-object basis.
 
Determinations of the SFR from ultraviolet (UV) \citep[e.g.][]{2007ApJ...670..928B} %(Bouwens+07)
and optical spectroscopy 
\citep[e.g., from the {\it Sloan} Digital Sky Survey, ][]{2011AJ....142...72E} 
are based on measurements of only around 10\% of the total integrated light, which escapes the dust absorption.  These must therefore be corrected upwards by large, uncertain extinction factors  (Fig. \ref{Fig1}). 
%The SFR density at $z$$>$2--3 is uncertain, since it is derived from UV surveys, which are highly affected by dust extinction. 
X-ray analyses of the BHAR, similarly, are prone to large uncertainties, because %they are based on incomplete (soft X-ray only) sampling of the total AGN flux. 
deriving the bolometric luminosity from the X-ray luminosity depends on uncertain bolometric corrections \citep[e.g.,][]{2009MNRAS.392.1124V} and on 
estimates of the Compton-thick population contribution \citep{2007ApJ...666...86F,2009ApJ...696..110T}. %(frontera+2007, treister+2009) 
%Rather than depending on so many model-dependent assumptions used to separate the stellar and 
%nuclear contribution to the measured total luminosities,  %from {\it Spitzer} and {\it Herschel} IR photometric data, 

IR emission line spectra can %measured from SPICA spectra will 
be used to {\it physically} separate through spectroscopy \citep[see, e.g.,][]{1992ApJ...399..504S} these contributions to the total integrated light on a galaxy-by-galaxy basis,  %SPICA will allow us to 
directly measure redshifts, SFRs, BHARs, metallicities and physical properties of gas and dust in galaxies. 
With SPICA, we will be able to do this for lookback times up to about 12 Gyrs. 
%The SPICA observations will enable us for the first time to obtain redshifts and to construct the SFR and BHAR functions (Fig. 1) 
%with measurements obtained in ''just one go" through spectroscopy. %, directly linked to the  properties of the galaxies. 

With its large, cold telescope and powerful instruments, SPICA will peer into the dust-enshrouded phases of galaxy formation and evolution, revealing the physical, dynamical and chemical states of the gas and dust.  
With such a unique power, SPICA will probe the histories of star-formation and black-hole accretion rates through cosmic time and how these processes drive galaxy evolution. It will provide detailed answers to the following questions:

\begin{enumerate}
%\item[$\bullet$] 
%\item[$\bullet$] $\bullet$ 
\item How does accretion and feedback from star formation and AGN shape galaxy evolution? (see section \ref{sect2.1})
%\item[$\bullet$] $\bullet$ 
\item How are metals and dust produced and destroyed in galaxies? What is the metallicity evolution in galaxies as a function of redshift? (see section \ref{sect2.2})
%\item[$\bullet$] $\bullet$ 
\item How and when do early black-holes and starburst appear close to the epoch of re-ionization? (see section \ref{sect2.3})
%\item[$\bullet$] $\bullet$ 
\item How did primordial gas clouds collapse into the first galaxies and black holes? (see section \ref{sect2.4})
%\end{itemize}$\bullet$ 
%\item What are the histories of star-formation and black-hole accretion rates through cosmic time and how do these two processes drive galaxy evolution? 
\end{enumerate}

%physics %, in particular above $z$$\sim$2 ?
In the following four sections we briefly address each of the first four questions, reserving detailed analysis of the role of SPICA in answering these to four companion articles: %for the detailed presentation of the above four themes %,  of galaxy evolution studies, 
%We then refer to other four specialised articles for the detailed presentation of the main themes of galaxy evolution studies with SPICA: 
\textrm{i)} the role of AGN feeding and AGN feedback in galaxy evolution  \citep{gon17}; %(Gonzalez-Alfonso et al. 2017, in prep.); 
\textrm{ii)} the chemical evolution of galaxies and the rise of metals and dust \citep{f-o17}; %(Fernandez-Ontiveros et al. 2017, in prep.); 
\textrm{iii)} dust obscured star-formation and accretion histories from re-ionization using SPICA unbiased photometric  \citep{gru17} and 
low spectral resolution spectroscopic surveys \citep{kan17}; %(Gruppioni et al. 2017, Kaneda+17) 
\textrm{iv)} the first stars and galaxies \citep{ega17}. %(Egami et al. 2017, in prep.). 
%\citep[][respectively]{gon17, f-o17, gru17, ega17}, while lWe leave detailed analysis of the final question to Sect. \ref{sect3} of this paper. 
We will detail in Sect. \ref{sect3} the power of using the many IR fine structure lines which fall in the wavelength range of the SPICA instrument for distant AGN and starbursts up to a redshift of {\it z}=4. 
%Finally we show in another companion article \citep{kan17} the power of unbiased low-resolution wide-field mid-IR spectroscopic surveys, to help revealing and understanding the histories of star-formation and black-hole accretion.

\subsection{AGN Feeding and Feedback in the context of Galaxy Evolution} \label{sect2.1}

The correlations between SMBH masses and their host galaxy properties \citep{1998AJ....115.2285M, 2000ApJ...539L...9F, 2000ApJ...539L..13G}, %(Magorrian et al 1998, Ferrarese \& Merritt 2000, Gebhardt+00), 
and the bimodality of the colour distribution of local galaxies 
\citep[e.g.,][]{2001AJ....122.1861S, 2004ApJ...600..681B, 2007ApJS..173..293W}, %(e.g. Strateva et al. 2001, Baldry et al. 2004, Wyder+07), 
suggest that the growth of BH and stellar mass are %intimately 
related throughout the lifetime of a galaxy.
In other words, feedback between SMBH and galactic star formation may be in part responsible for the $M_{BH}$--$\sigma$  relationship seen in the local Universe
%This process may lead to the M$_{\rm BH}$--$\sigma$ relationship seen in the local Universe 
\citep[e.g.,][]{1998A&A...331L...1S, 2005Natur.433..604D, 2005MNRAS.361..776S}. %(e. g., Silk \& Rees 1998; di Matteo et al. 2005; Springel et al. 2005). 
It is precisely this feedback on the dense, circum-nuclear ISM that we can study with SPICA in the far-IR.
%For all the details of this study and the assessment of the SPICA observations with detailed simulations, we refer to the companion paper by \citet{gon17}, %Gonz\'alez-Alfonso et al (2017), 
%while briefly introducing here this topic.

% The  from
{\it Herschel} spectroscopic observations of far-IR OH lines \citep{2010A&A...518L..41F, 2011ApJ...733L..16S, 2014A&A...561A..27G, 2017ApJ...836...11G} %, that 
%%(Fischer et al. 2010, Sturm et al. 2011, Gonz\'alez-Alfonso et al. 2014, Gonz\'alez-Alfonso et al. 2017).has %studies have 
have shown that fast, molecular outflows are common among AGN-powered local ULIRGs.  SPICA spectroscopy will allow us to search for these outflows well beyond the nearest galaxies, reaching up to $z$=1--2, and therefore providing an estimate of the impact of AGN-driven feedback at the peak epoch of star formation.
With the SAFARI instrument on SPICA we will be able to detect the spectral signatures of outflowing dense molecular gas (through P-Cygni profiles on the OH 79\,$\mu$m and 119\,$\mu$m lines and blue-shifted high velocity wings in the OH 65\,$\mu$m line) in ULIRGs like Mrk 231 out to $z$=1.5 in a few hours of integration.  
This will enable surveys of hundreds of galaxies at these redshifts, providing a measure of the
demographics of molecular feedback in IR-luminous galaxies at the peak of the SFR density.

%Although blue-shifted absorption is a direct signature of outflowing molecular gas and feedback on the star-forming ISM, 
Far-IR spectroscopy %with SPICA 
can also provide direct evidence for the feeding of the SMBHs and central starbursts.  Inflowing gas can be identified through inverse P-Cygni profiles or 
redshifted absorption wings of OH and [OI]63\,$\mu$m, as shown by {\it Herschel} %, as seen, e.g., in the galaxies NGC4418 and Zw049 by {\it Herschel} %(see Fig. 7b;  
\citep{2012A&A...541A...4G, 2015A&A...580A..52F}. %Eduardo+2012, Falstad et al. 2015).  
%[see Fig. \ref{Fig7}, right panel;][]
%Sensitive far-IR spectroscopy can in this way provide a complete and unique An IR space telescope such 
SPICA will therefore be able to measure the dynamics of the molecular gas in and around the nuclei of 
rapidly evolving, dusty galaxies and %.SPICA will 
study both AGN accretion and energetic feedback in significant samples of dusty galaxies over the past 10 Gyr.
For a more detailed discussion of the capabilities of SPICA to detect galaxy feedback, see the companion paper by  \citet{gon17}. 
%Gonzalez-Alfonso et al. (2017).

\subsection{The Rise of Metals and Dust} \label{sect2.2}

%The evolution of galaxies is %deeply 
%linked to their chemical evolution, which reflects into metallicity and dust evolution. 
Galactic evolution is intimately tied to the production of metals and dust. 
Elements heavier than He (i.e. $\textquoteleft$$\textquoteleft$metals'') play a major role in gas cooling and, as a consequence, they are critical in determining the conditions of cloud collapse and ultimately of star and planet formation. % of stars and planets. 
The metallicity in galaxies is determined by the cumulative effects of star formation, outflows, accretion and the radial redistribution of matter. 
Traditional metallicity diagnostics, based on %bright 
optical lines, are %however %strongly 
biased towards dust-free regions, % and they critically depend on the assumed electron temperatures. Often this approach 
yielding %in dust-obscured galaxies 
metallicities significantly lower than those inferred from the dust mass or IR lines %especially %, infrared-bright 
\citep[e.g.,][]{2017MNRAS.470.1218P, 2010A&A...518L.154S, 2013ApJ...777...96C}. %(e.g. Santini et al. 2010; Croxall+2013; miguel et al. 2017). 
Furthermore, %because of temperature inhomogeneities in galaxies, %different regions of a galaxy have different temperature,
%optically derived abundances are influenced by %uncertainties and 
%temperature fluctuations, %of the electron , because of the different %electron temperatures in 
 %, where metals originate, have . These fluctuations 
%which do not affect the IR lines, which %because they %se latter 
%originate from levels %so 
%close to the ground level %, that their population does not depend on the electron temperature
%to populate them, the electron temperature does not play an important role 
temperature variations within galaxies, which can strongly influence optically derived abundances, have little effect on IR-derived values, 
which are determined from lines lying close to the ground state \citep{2001A&A...367..949B}. %(Bernard-Salas et al. 2001). Rest-frame 
%Mid- and far-
IR fine-structure lines observed with SPICA 
provide extinction-free measurements of the metallicity of galaxies, nearly independent 
of ionization, density and temperature out to redshift $z$$\sim$3. %--4. % (Fig.~8). 
Moreover, SPICA can detect mid-IR Hydrogen recombination lines in galaxies at intermediate redshifts ($z$$\sim$1.5--2), %as well, 
which, together % in combination 
with the forbidden lines, allow for direct determination of the %Neon, Sulfur, Nitrogen, Oxygen and Iron 
Ne, S, N, O and Fe abundances.
%We refer to %the companion paper by 
%\citet{f-o17} for the details of this study and the assessment of the SPICA observations. %Fern\'andez-Ontiveros et al (2017).

Dust also plays a critical role in heating and cooling the ISM in galaxies. It forms in the dense, enriched atmospheres of evolved stars, novae, supernovae and %in 
dense molecular clouds \citep{2009MNRAS.397.1661V, 2013A&A...555A..99Z, 2015MNRAS.454.4250M, 2016A&A...587A.157B} %(e.g. Valiante et al. 2009, Zhukovska \& Henning 2013, Marassi et al. 2015, Bocchio et al. 2016)
and is %t is %easily 
destroyed by shocks, %as well as 
sputtering and intense radiation fields. However, its dominant formation and 
destruction channels in different environments are %still 
poorly understood %known %a matter of great debate 
\citep{2015PKAS...30..283K, 2011Sci...333.1258M, 2016A&A...590A..65M}. %(e.g. Kemper 2015; Matsuura et al. 2011; Micelotta et al. 2016). 
%SPICA will 
%It provides the shielding that allows dense gas to turn molecular, as well as the surfaces for H$_2$ formation. Through the photoelectric effect, small grains 
%provide most gas heating in PDRs. 
%Because dust , 
IR spectroscopy can uniquely measure the %mass of 
dust mass produced by evolved stars and supernovae in nearby galaxies, allowing a detailed study of the dust mineralogy and composition, 
via the mid-IR {\it SiO$_2$}, {\it FeO}, {\it FeS} and crystalline silicate features \citep[see, e.g.,][]{2006ApJ...638..759S, 2008ApJ...673..271R}. 
% (Spoon+2006, rho+2008) %With its spectral and continuum capability, 
SPICA will trace the abundance and evolution 
of the dust components within galaxies, constraining the local conditions, such as ionization, radiation field, dust structures and overall dust-to-gas ratios, 
giving us clues to the chemical evolution of galaxies \citep{2012ApJ...744...20S, 2014A&A...563A..31R}. 
%(e.g. Galliano et al. 2008; Sandstrom et al 2012; Remy-Ruyer et al. 2014).
For the details of these investigations and the assessment of the SPICA observations, we refer to the companion papers by \citet{flo17} and \citet{f-o17}, for local and distant galaxies, respectively. %van der Tak et al (2017).

\subsection{Towards the Epoch of Re-ionization: early Black Holes and Starbursts} \label{sect2.3}

The deepest cosmological surveys with {\it Herschel} \citep[e.g.,][]{2012MNRAS.424.1614O, 2013MNRAS.432...23G, 2013A&A...553A.132M, 2014ARA&A..52..373L} 
%(e.g. Oliver et al. 2012; Gruppioni et al. 2013; Magnelli et al. 2013; Lutz 2014) 
mapped out the SFR %star-formation rate 
density to $z$$\sim$3 for the first time, but detected only small numbers of the most luminous 
($L\gtrsim$10$^{12}$L$_{\odot}$) star-forming galaxies at $z$$>$3. High redshift ($z$$>$6) quasars have been detected, containing 
black holes as massive as 10$^{10}$M$_{\odot}$ \citep{2015Natur.518..512W, 2015ApJ...806..109J}, %(Wu et al., 2015, Jun et al 2015)
but their origin, demographics and role in re-ionization are still unclear. 
Deep SPICA/SMI photometric surveys will extend the study of the  BHAR %black-hole accretion rate 
and  SFR %the star-formation rate 
density well beyond $z$$\sim$3, detecting, in the 34\,$\mu$m observed wavelength, the hot dust around high-redshift QSOs, as well as starburst-dominated galaxies at redshifts out to $z$$\sim$6. % (Fig.9). 
%Fig. \ref{Fig8} (left) shows the redshift-luminosity space covered by six 10'$\times$10' frames of SMI (0.2 deg$^2$) at the confusion limit of 5$\mu$Jy. 
%Fig. \ref{Fig8} (right) also shows the detailed SED from radiative transfer models \citep{2009A&A...502..541E} %(Efstathiou & Siebenmorgen 2009) 
%fitting the $z$=4 hyperluminous galaxy GN20 \citep{2009ApJ...694.1517D}. %(daddi+2009)
Due to its large field of view of 10$^{\prime}\times12^{\prime}$, 
SMI will map large sky areas to the confusion limit (around 5\,$\mu$Jy) in relatively short times, with an 
effective surveying speed hundred times faster than JWST.\footnote{The JWST Design Reference Mission includes a MIRI (16\,$\mu$Jy, 5\,$\sigma$) 10$^{\prime}$$\times$9$^{\prime}$ survey at 21\,$\mu$m. With the MIRI field of view of 1.25$^{\prime}$$\times$1.88$^{\prime}$, this survey will need 160hrs. A similar SMI survey at 34\,$\mu$m, would take 1 hr (12\,$\mu$Jy, 5\,$\sigma$). For detecting mid-IR sources SPICA will be over 100 times faster than JWST.} 
%For this reason  SMI will be the best instrument for discovering high-$z$ AGN and starburst-dominated galaxies in large deep surveys. 
%Together with the earliest X-ray luminous AGN pinpointed by {\it Athena}, SPICA will also provide the full cosmic history of accretion (see sect. \ref{sect2.1.1}). 
%Follow-up observations with the next generation of ELTs and ALMA %, as well as with the SPICA spectrographs, 
%will fully characterize the faint, red galaxies detected by SMI up to $z$$\sim$5--6. These surveys will focus both on blank fields and on known proto-clusters. The latter will 
Deep and wide photometric surveys with SPICA will 
allow us to study the build-up of the progenitors of the elliptical galaxies dominating local galaxy clusters, and thus to probe environmental-dependent evolution \citep[e.g.,][]{2014A&A...570A..55D, 2016MNRAS.461.1719C}.
For a more detailed description of potential photometric surveys with SPICA, and their role in extending our knowledge of the population of IR-bright galaxies at $z$$>$3, we refer to the companion paper by \citet{gru17}. %Gruppioni et al (2017).

\subsection{The First Stars and Galaxies} \label{sect2.4}

Through chemical enrichment and the production of dust, the earliest stars imprint their signature on the ISM of high-redshift galaxies.
The first generation of quiescent, self-gravitating primordial or metal-poor gas clouds may be too faint to be detected directly with SPICA \citep[e.g.,][]{2005PASJ...57..951M, 2006ApJ...643...26S, 2013ApJ...768..130G}. 
%(e.g. Mizusawa et.al. 2005; Gong et.al. 2013, Santoro & Shull 2006). 
However, during the hierarchical structure formation process, both the merging of primordial clouds and feedback due to supernovae and stellar winds can 
produce pockets of shock-heated gas. 
%In Fig. \ref{Fig11}, the gas density (left panel), gas temperature (middle panel) and H$_2$ surface density 
%(right panel) of a simulated $z=6$ galaxy are shown \citep[][see also \citealt{gallerani:2016}]{2017MNRAS.465.2540P}. %(Pallottini+16)
It is postulated that molecular clouds can be efficiently formed in the cooling gas behind these shocks %produced during the blowaway 
\citep{ferrara:1998, ciardi:2001}. 
%and if these clouds are warm and massive enough, they could be detected with SAFARI through their H$_2$ emission.
If these {\it H$_2$} 
clouds are warm and massive enough,  
(e.g., 10$^{10}${\it M}$_{\odot}$ of {\it T}=200 {\it K} gas and 10$^{8}${\it M}$_{\odot}$ of {\it T}=1000 {\it K} gas), 
they are detectable in their shock-excited {\it H$_2$}  line emission with SAFARI.
Such systems of warm massive  {\it H$_2$} gas reservoirs are already 
known at lower redshift \citep{2006ApJ...652L..21E, 2012ApJ...751...13O}, %(Egami et al. 2006; Ogle et al. 2012), 
but can be detected by SPICA out to $z$$\sim$10. % (Fig. \ref{Fig10}, red lines). 

%SPICA's 
The ultimate challenge for %of a far-IR space telescope such 
SPICA will be to catch a glimpse of the first (i.e., Pop III) galaxies. 
Theoretical models predict that Pop III star clusters should produce a 
large amount of dust quickly, as massive pair instability supernovae (PISNe) explode \citep[e.g.,][]{2004MNRAS.351.1379S}. 
%(e.g., Schneider et al. 2004). 
When the dust is released and heated by hot, main sequence Pop III stars, the resultant mid-IR spectra %will 
exhibit strong quartz ({\it SiO$_2$}) emission features \citep{sch17}, 
which SPICA may detect. % (Fig. \ref{Fig12}). 
This will give a rare glimpse into the dust-production mechanism of Pop III stars and %, % . These results suggest that SPICA will 
provide strong constraints for %existing Pop III 
their evolutionary models. %of the first stars. 
%For the details of this study and the predictions for SPICA observations, we refer to the companion paper by 
For a detailed discussion of the potential for SPICA to detect the cooling signatures of young, luminous galaxies at high redshift, 
see the companion paper by \citet{ega17}. %Egami et al (2017).

\begin{table}[h]
\caption{For the top nine most abundant metals in the universe, this table gives the fine-structure lines that will be observed by SPICA in the redshift range $0 < z < 4$. I.P. denotes the ionization potential of each transition. Solar abundances \citep[written as $\log \it{X} / \it{H} + 12$, from][]{2010Ap&SS.328..179G} 
 and critical densities for collisional de-excitation have been taken from \citet{1985ApJ...291..722T, 1992gim..conf..275G, 1993ApJS...88...23G}. 
%(Tielens & Hollembach 1985, Genzel 1992, Greenhouse et al 1993)
} \label{fslines}
\centering
\begin{small}
\begin{tabular}{lclccccccccccc}
\hline \hline
\\[-0.2cm]  
Atom & Solar & Ion/ & $\lambda$ & I.P. & {\it n}$_{\rm crit}$\\
&  abundance &  Line & ($\rm{\mu m}$) & (eV) & (cm$^{-3}$)\\[0.1cm]
\hline \\[-0.2cm]
%O & 8.69 \\
O & 8.69 & [OI] & 63.18 & -- & 4.7$\times$10$^{5}$ \\
 &  & [OI] & 145.5 & -- & 9.5$\times$10$^{4}$ \\
 &  & [OIII] & 51.81 & 35.12 & 3.6$\times$10$^{3}$ \\
 &  & [OIII] & 88.36 & 35.12 & 5.1$\times$10$^{2}$ \\
 &  & [OIV] & 25.89 & 54.93 & 1.0$\times$10$^{4}$ \\
%C & 8.55 \\
C & 8.55 & [CII] & 157.7 & 11.26 & 2.8$\times$10$^{3}$ \\
%Ne & 7.93 \\
Ne & 7.93 & [NeII]  & 12.81 & 21.56 & 5.4$\times$10$^{5}$ \\
 &  & [NeIII] & 15.56 & 40.96 & 2.9$\times$10$^{5}$ \\
 &  & [NeIII] & 36.01 & 40.96 & 4.2$\times$10$^{4}$ \\
 &  & [NeV]   & 14.32 & 97.12 & 3.8$\times$10$^{5}$ \\
 &  & [NeV]   & 24.32 & 97.12 & 5.4$\times$10$^{4}$ \\
 &  & [NeVI]  &  7.65 & 126.21 & 2.5$\times$10$^{5}$  \\
%N & 7.83 \\
N & 7.83  & [NII] & 121.9 & 14.53 & 3.1$\times$10$^{2}$ \\
 &  & [NII] & 205.2 & 14.53 & 4.8$\times$10$^{1}$ \\
 &  & [NIII] & 57.32 & 29.60 & 3.0$\times$10$^{3}$ \\
%Mg & 7.60 \\
Mg & 7.60 &  [MgIV] &  4.49 & 80.14 & 6.3$\times$10$^{6}$ \\
 &  & [MgV] &  5.61 & 109.24 & 2.0$\times$10$^{6}$ \\
%  &  & [MgV] & 13.52 & 109.24 \\
%  &  & [MgVII] &  5.50 & 186.51 \\
%  &  & [MgVII] &  9.01 & 186.51 \\
%  &  & [MgVIII] &  3.03 & 224.95 \\
%Si & 7.51 \\
%  &  & [SII] & 68.47 & 0.00 \\
%  &  & [SII] & 129.7 & 0.00 \\
Si & 7.51  & [SiII] & 34.82 & 8.15 & 3.4$\times$10$^{5}$ \\
%  &  & [SIIx] &  2.58 & 303.17 \\
%  &  & [SIIx] &  3.94 & 303.17 \\
%  &  & [SIVI] &  1.96 & 166.77 \\
%  &  & [SIVII] &  2.48 & 205.05 \\
  &  & [SiVII] &  6.49 & 205.05 &  5.0$\times$10$^{6}$ \\
%Fe & 7.50 \\
Fe & 7.50 & [FeI] & 24.04 & -- & 3.1$\times$10$^{6}$ \\
  &  & [FeI] & 34.71 & -- & 3.0$\times$10$^{6}$ \\
%  &  & [FeI] & 54.31 & 0.00 \\
%  &  & [FeI] & 111.2 & 0.00 \\
%  &  & [FeII] &  4.08 & 7.90 \\
%  &  & [FeII] &  4.08 & 7.90 \\
%  &  & [FeII] &  4.11 & 7.90 \\
%  &  & [FeII] &  4.43 & 7.90 \\
%  &  & [FeII] &  4.61 & 7.90 \\
%  &  & [FeII] &  4.67 & 7.90 \\
%  &  & [FeII] &  4.89 & 7.90 \\
%  &  & [FeII] &  5.06 & 7.90 \\
%  &  & [FeII] &  5.34 & 7.90 \\
%  &  & [FeII] &  5.67 & 7.90 \\
% &  & [FeII] &  6.72 & 7.90 \\
% &  & [FeII] & 17.94 & 7.90 \\
%  &  & [FeII] & 24.52 & 7.90 \\
 &  & [FeII] & 25.99 & 7.90 & 2.2$\times$10$^{6}$ \\
 &  & [FeII] & 35.35 & 7.90 & 3.3$\times$10$^{6}$ \\
%  &  & [FeII] & 35.78 & 7.90 \\
%  &  & [FeII] & 51.30 & 7.90 \\
%  &  & [FeII] & 87.38 & 7.90 \\
% &  & [FeIII] & 22.93 & 16.19 \\
%  &  & [FeIII] & 33.04 & 16.19 \\
%  &  & [FeIII] & 51.68 & 16.19 \\
%  &  & [FeIII] & 105.4 & 16.19 \\
%  &  & [FeVII] &  7.81 & 99.10 \\
%  &  & [FeVII] &  9.53 & 99.10 \\
%S & 7.12 \\
%  &  & [SI] & 25.25 & 0.00 \\
%  &  & [SI] & 56.31 & 0.00 \\
S & 7.12 & [SIII] & 18.71 & 23.34 & 1.7$\times$10$^{4}$ \\
 &  & [SIII] & 33.48 & 23.34 & 2.0$\times$10$^{3}$ \\
 &  & [SIV]  & 10.51 & 34.79 & 5.6$\times$10$^{4}$ \\
% Al & 6.45 \\
%  &  & [AlI] & 89.24 & 0.00 \\
%  &  & [AlV] &  2.91 & 120.00 \\
%  &  & [AlVI] &  3.66 & 153.83 \\
%  &  & [AlVI] &  9.12 & 153.83 \\
%  &  & [AlVIII] &  3.69 & 241.44 \\
%  &  & [AlVIII] &  5.85 & 241.44 \\
%Ar & 6.40 \\
Ar & 6.40 & [ArII] &  6.99 & 15.76 & 1.9$\times$10$^{5}$ \\
 &  & [ArIII] &  8.99 & 27.63 & 3.1$\times$10$^{5}$ \\
 &  & [ArIII] & 21.83 & 27.63 & 3.5$\times$10$^{4}$ \\

\hline
\end{tabular}
\end{small}
%\tablefoot{
%Solar abundances as log X\slash H+12 from Grevesse+2010 Ap\&SS, 328: 179-183. Transitions from http://www.mpe-garching.mpg.de/iso/linelists/FSlines.html
%}
\end{table}

\begin{figure*}[ht]
\begin{center}
  \begin{minipage}[c]{0.7\textwidth}
    \centering
    \includegraphics[width=\textwidth]{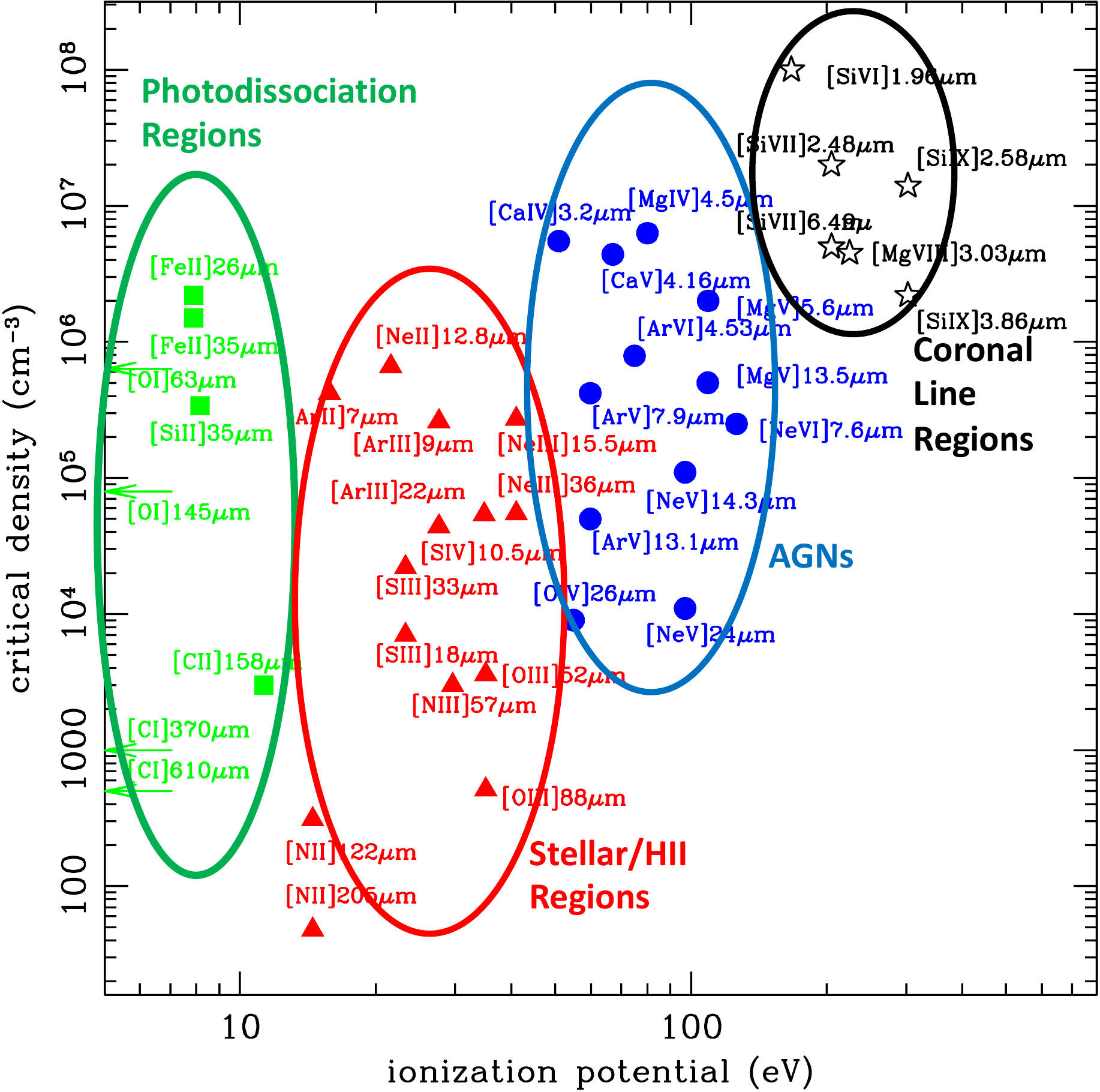}
  \end{minipage}\hfill
  \begin{minipage}{0.25\textwidth}
    \caption{Critical densities for collisional de-excitation versus the ionization potential of the IR fine-structure lines \citep[adapted from][]{1992ApJ...399..504S}.}\label{Fig2}
   \vspace{-6.3cm} %(Spinoglio \& Malkan 1992).
  \end{minipage}\hfill
%\includegraphics[width=0.8\textwidth]{ion_density_large.pdf}
%%\includegraphics{Figure1.pdf}
%\caption{Critical densities for collisional de-excitation versus the ionization potential of the IR fine-structure lines \citep[adapted from][]{1992ApJ...399..504S}. %(Spinoglio \& Malkan 1992). 
%SPICA, with its sensitivity and broad wavelength coverage, will provide access to key mid- and far-IR lines probing a wide range of ionization, density and ISM phases.
%}\label{Fig2}
\end{center}
\end{figure*}

\section{Infrared Spectroscopic Probes of Star Formation and Black Hole Accretion} \label{sect3}

To understand galaxy evolution, we need to measure the rate at which stars form and black holes accrete matter as a function of time. 
The apparently similar shapes of the histories of star formation and black hole accretion with redshift (see Fig. \ref{Fig1}), the fact that most AGN also display enhanced star formation, 
%<\citep[e.g.,][]{2003MNRAS.346.1055K} > 
along with the black hole-stellar mass correlation seen in the local Universe \citep{1998AJ....115.2285M, 2000ApJ...539L...9F, 2000ApJ...539L..13G}, %(Magorrian et al 1998, Ferrarese \& Merritt 2000, Gebhardt+00), 
all suggest that the two processes are physically linked.

The mid- to far-IR spectral range includes a suite of atomic and molecular lines and features, covering a wide range of physical conditions (excitation, density, ionization, radiation field, metallicity and dust composition) in galaxies  \citep[Fig. \ref{Fig2}, and][]{1992ApJ...399..504S}. %(Spinoglio \& Malkan 1992).
Table\,\ref{fslines} lists the set of atomic and ionic fine structure lines that will be covered by the SMI and SAFARI spectrometers onboard SPICA, for the top nine most abundant elements in the universe after H and He (the so-called $\textquoteleft$$\textquoteleft$metals''). 
These lines reveal the detailed physics in the various phases of the interstellar medium (ISM), from  HII and photo-dissociation regions (PDR), 
to the Narrow Line Regions %(NLR) %and Coronal Line Regions (CLR) 
excited by AGN. In the highly opaque, dust obscured ISM of actively star-forming galaxies and AGN, 
the IR lines are among the few probes of 
the physical conditions in the gas and dust clouds surrounding the SMBH or young, hot stars. 

The neutral gas surrounding star-forming regions can be traced using the temperature sensitive [OI] lines at 63\,$\mu$m and 145\,$\mu$m, 
%that \textit{Herschel} was not sensitive enough to map, 
while the ionised gas can be studied using many different tracers to measure temperature, density and abundances. The %{\bf 
[NII]122/205\,$\mu$m, [OIII]52/88\,$\mu$m, [SIII]18.7/33.5\,$\mu$m, and [NeIII]15.6/36.0\,$\mu$m line ratios are individually sensitive to the density; since the electron temperatures are generally higher than the excitation states of the lines connecting these line ratios, they are not sensitive to the temperature. %} %They all sample the same element in the same ionization state. 

The strength and hardness of the radiation field can be derived from line pairs of the same element in different ionization states and here again the SPICA wavelength range is ideal. Examples are the [NIII]57\,$\mu$m to [NII]122 or 205\,$\mu$m ratio, the [NeIII]15.6\,$\mu$m to [NeII]12.8\,$\mu$m ratio and the [OIV]26\,$\mu$m to [OIII]52 or 88\,$\mu$m ratio. The ratio of the [OIII] to [NIII] lines can be used to derive relative abundances of O and N, two key elements in the gas chemistry \citep[see][]{f-o17}. Other strong lines, such as [SiII]34.8\,$\mu$m or [CII]158\,$\mu$m, probe the interface between the neutral and ionized gas. 

\begin{figure*}[h]
\begin{center}
\includegraphics[width=0.432\textwidth]{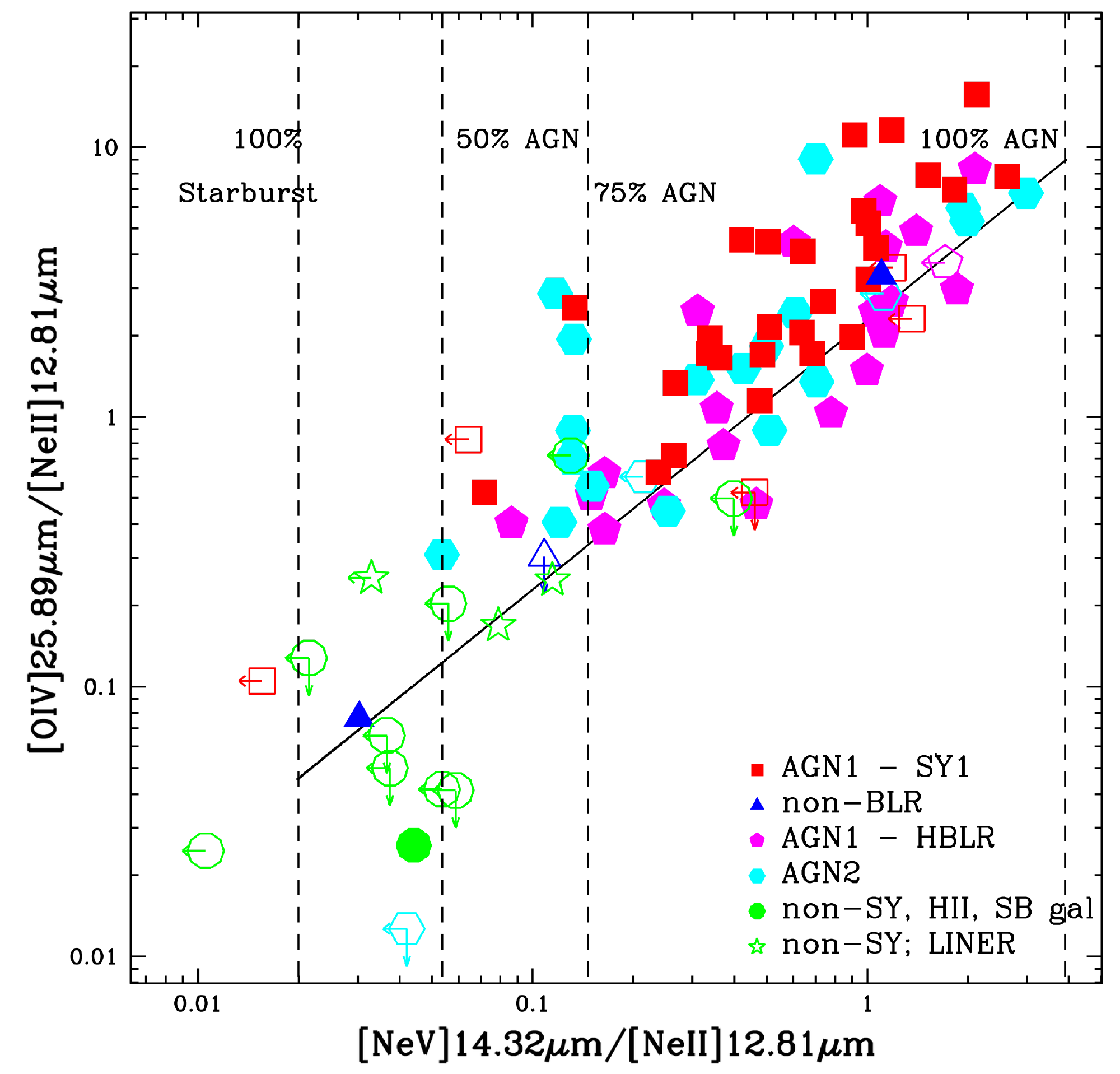}~
\includegraphics[width=0.405\textwidth]{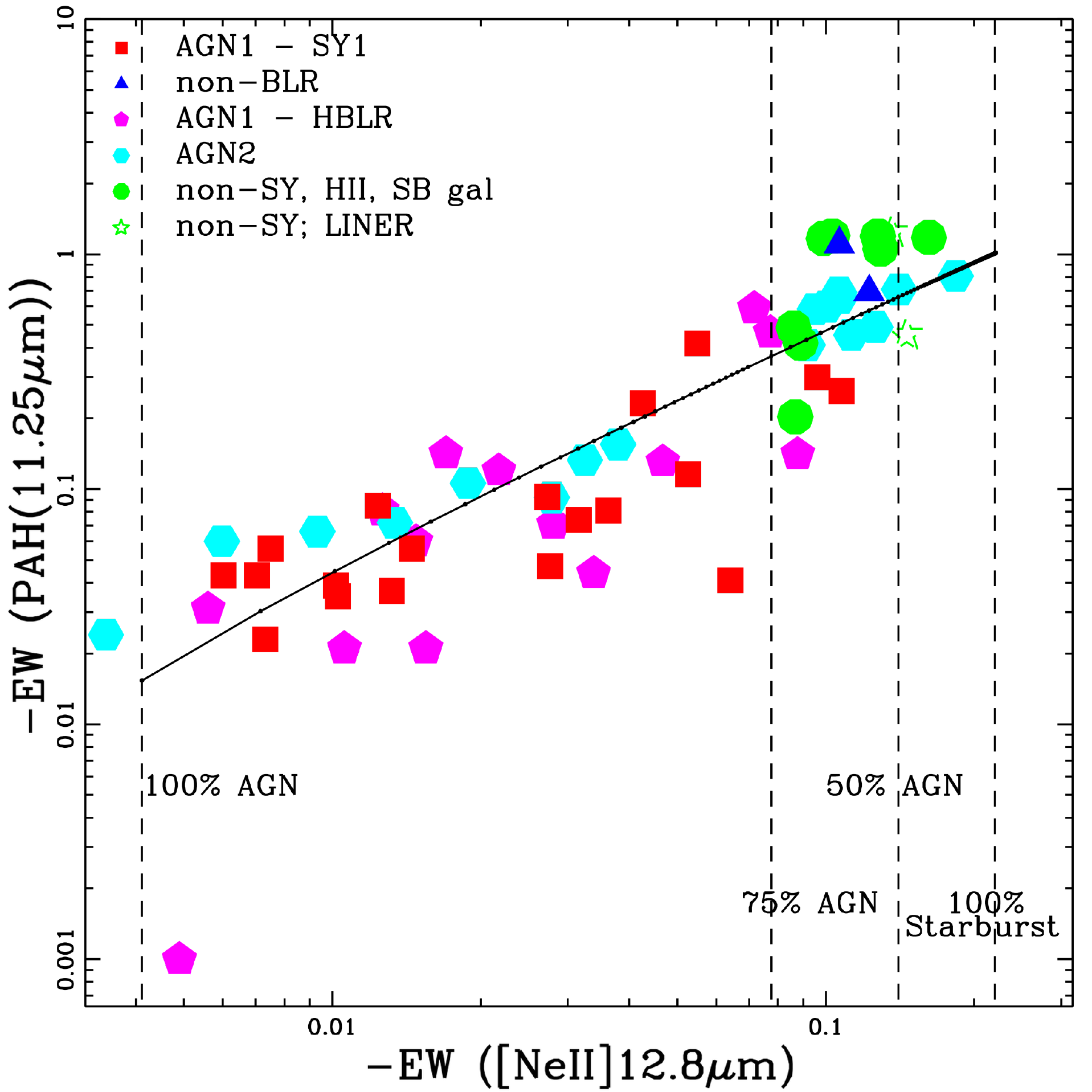}
\caption{{\it Left panel: }[NeV]14.3\,$\mu$m/[NeII]12.8\,$\mu$m line ratio versus [OIV]25.9\,$\mu$m/[NeII]12.8\,$\mu$m line ratio. Both axes
correlate with the strength of the AGN.  
The black line shows the behaviour of the analytical model \citep{2010ApJ...709.1257T}. %Tommasin+2010.
{\it Right panel: } [NeII]12.8\,$\mu$m equivalent width versus PAH 11.25\,$\mu$m equivalent width.  Both quantities in this plot correlate with the strength of the
star-formation component in each galaxy. The black line shows the behaviour of the analytical model \citep{2010ApJ...709.1257T}. }%Tommasin+2010.}. 
\label{Fig3}
\end{center}
\end{figure*}

The SFR can be obtained via low-ionization lines (e.g. [NeII]12.8\,$\mu$m, [NeIII]15.5\,$\mu$m, [SIII]18.7\,$\mu$m), while the AGN accretion rate can be measured via high-ionization lines (e.g., [OIV]25.9\,$\mu$m, [NeV]14.3\,$\mu$m and 24.3\,$\mu$m). 
To disentangle emission from an AGN and that of star formation 
\citep{2005ApJ...623..123S, 2006ApJ...640..204A,  2007ApJ...658..314H, 2008ApJ...679..101S, 2012ApJ...744....2A, 2014ApJ...790..124S, 2015ApJ...799...21S, 2016ApJS..226...19F}, studies have exploited fine-structure lines and the characterization of the broad emission features arising from polycyclic aromatic hydrocarbons (PAHs) and the mid-IR thermal dust continuum
\citep{1998ApJ...498..579G,  2007ApJ...656..148A, 2007ApJ...667..149F, 2007ApJ...654L..49S, 2008ApJ...684..853L,  2008ApJ...676..836T, 2009ApJ...700L.149V, 2010ApJ...709.1257T, 2011ApJ...730...28P}. 
%(Spinoglio et al. 2005; Lutz et al. 2008; Schweitzer et al. 2008; Tommasin et al. 2008, 2010; Alonso-Herrero et al. (2012); Spinoglio et al. 2015; Fernandez-Ontiveros et al. 2016).  
%To give and example of how the IR fine-structure lines can be used together with the dust emission features, 
%{\bf 
In Fig. \ref{Fig3} we show the line ratios of [OIV]25.9\,$\mu$m/[NeII]12.8\,$\mu$m versus [NeV]14.3\,$\mu$m/[NeII]12.8\,$\mu$m, for a sample of local active galaxies. Both these ratios measure the strength of the AGN. We also show the equivalent width of the PAH emission feature at 11.25\,$\mu$m, as a function of the equivalent width of [NeII]12.8\,$\mu$m, which measure the strength of the starburst component in galaxies \citep{2010ApJ...709.1257T}. %}
%versus the equivalent width of PAH %is shown 
% that illustrates how this diagnostic diagram can measure the dominance of AGN versus starburst power in galaxies 

\begin{figure*}[h]
\begin{center}
\includegraphics[width=0.8\textwidth]{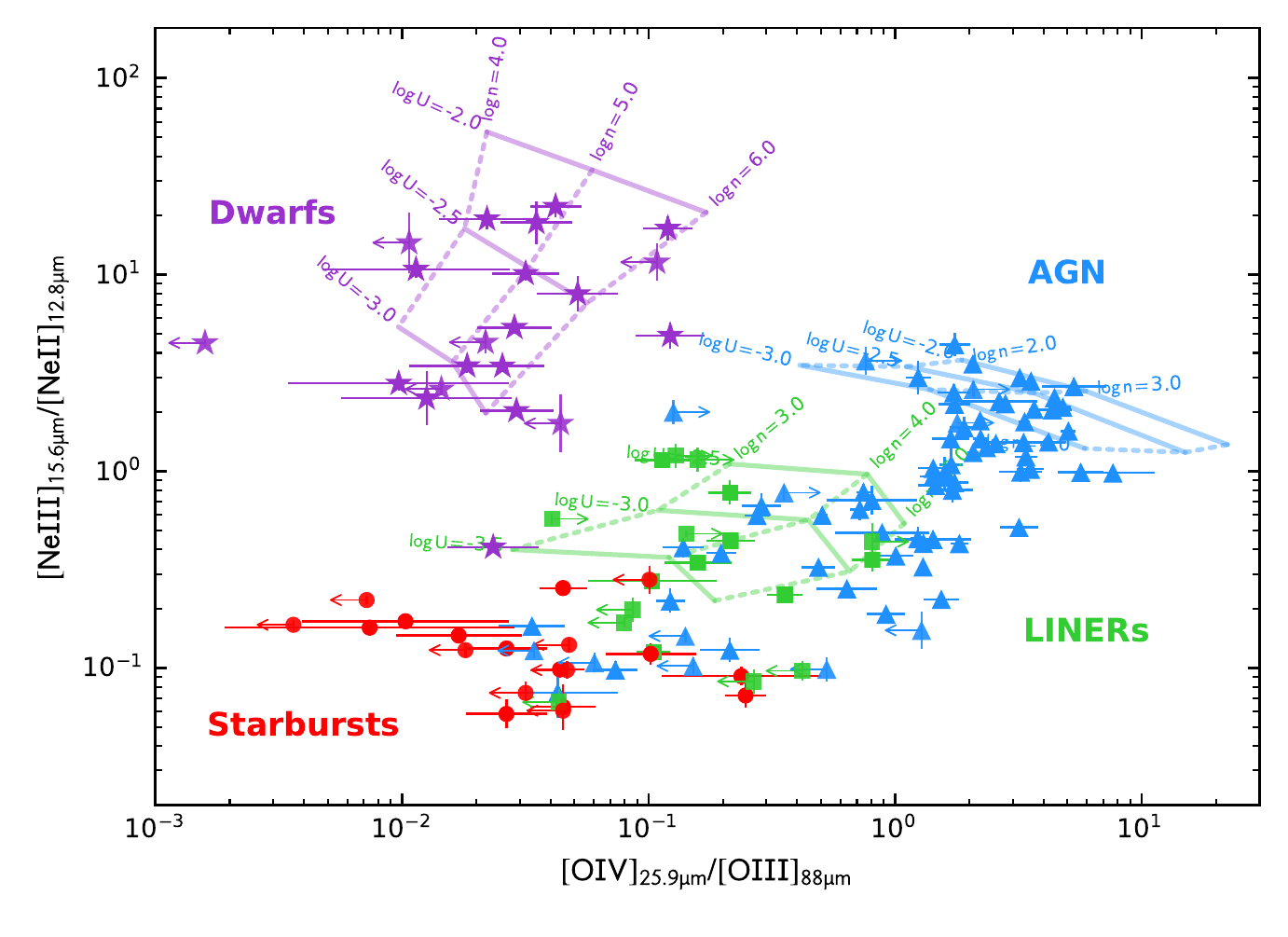}
\caption{Line ratio diagram of [NeIII]15.6\,$\mu$m/[NeII]12.8\,$\mu$m versus [OIV]26\,$\mu$m/[OIII]88\,$\mu$m, showing the comparison of the data for local Universe AGN, LINERs, starburst galaxies, and dwarf galaxies with with models including a range of the ionization parameters U and number densities n.
% Current starburst models are omitted because they fail to reproduce the data 
This diagram is able to disentangle nuclear activity from AGN, from normal star formation, as well as from low metallicity star formation, in e.g. dwarf galaxies \citep{2016ApJS..226...19F}. %(Fernandez-Ontiveros et al. 2016, ApJ, in press).
SPICA will be able to measure these line ratios and separate the power sources in active galaxies out to $z$$\sim$4.
}\label{Fig4}
\end{center}
\end{figure*}

\begin{figure}
\begin{center}
\includegraphics[width=\columnwidth]{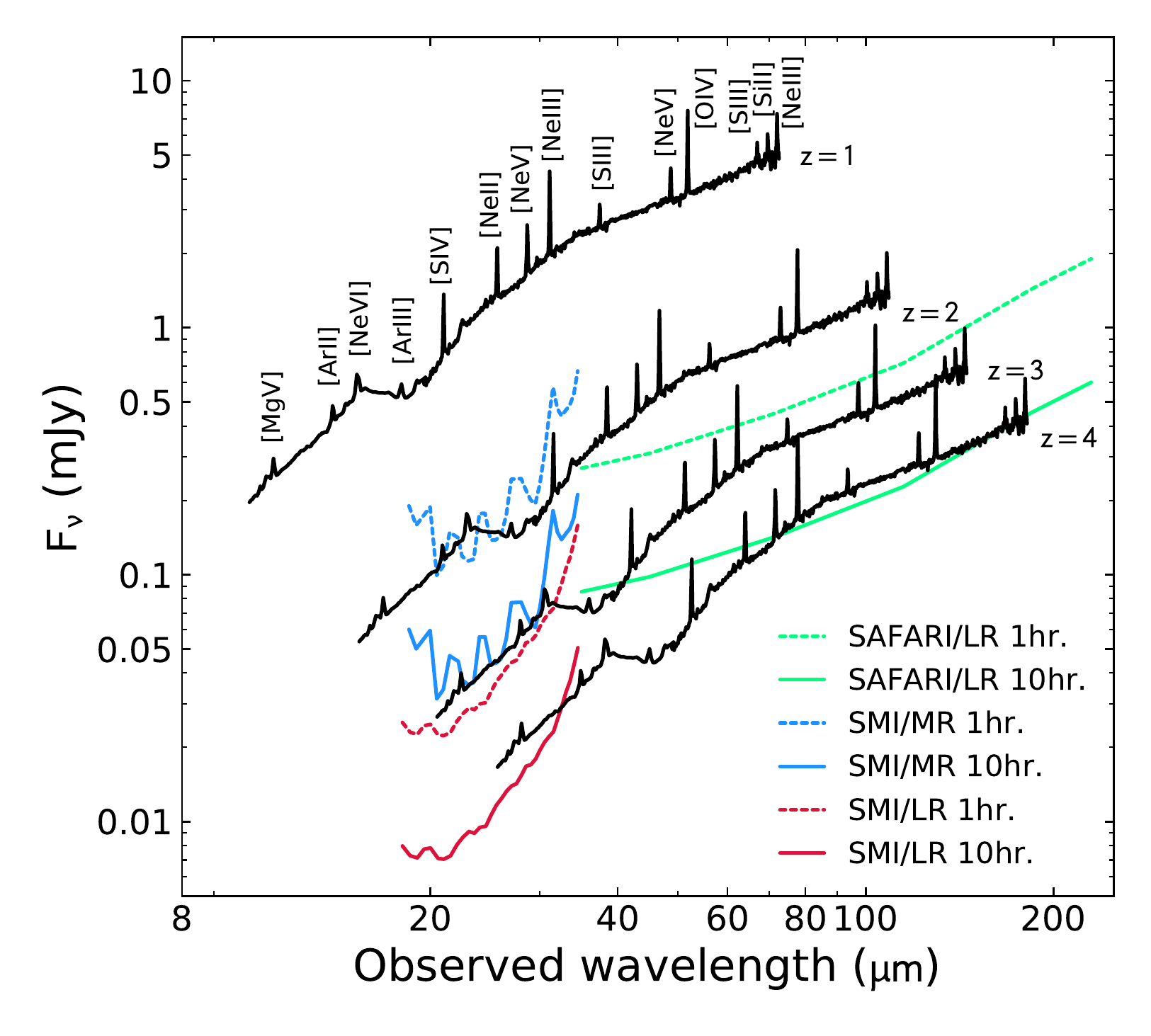}
\caption{IR spectrum of MCG-3-34-64, a nearby active galaxy, rescaled to a luminosity of {\it L}=10$^{12}$L$_{\odot}$ at redshifts from 1 to 4. At $z$=3, the knee luminosity is {\it L*}=10$^{12}$L$_{\odot}$, implying that SPICA will be able to map the bulk of the galaxy population up to this redshift. The 5$\sigma$ sensitivities of SAFARI (in low resolution mode, LR) and SMI (in medium and low resolution, MR and LR) are shown for integration times of 1 hr and 10 hrs.
}\label{Fig5}
\end{center}
\end{figure}

Where {\it Spitzer} and {\it Herschel} detected only the brightest galaxies at redshifts $z$$>$0.5  \citep{2007ApJ...660.1060V, 2008ApJ...675.1171P, 2009ApJ...699..667M, 2010A&A...518L..36S,  2014ApJ...786...31R, 2016ApJ...824..146Z}, %(Valiante+07, Sturm et al. 2010, Pope et al. 2008; Menendez-Delmestre et al. 2009; Riechers et al. 2014, Zhao, Y. et al 2016), 
SPICA will reveal the properties of the bulk of the IR galaxy population up to redshifts of $z$$\simeq$4 %, as it has been assessed in the study on the predictions of the IR line luminosity functions from IR spectroscopic cosmological surveys from 
\citep{2012ApJ...745..171S}. The detected lines and dust features will allow us to unambiguously quantify the contribution of star formation and AGN 
to the bolometric luminosity in each galaxy.  In Fig. \ref{Fig4} we illustrate how
%to its bolometric luminosity %(e.g., via diagrams such as the one given in Fig. \ref{Fig3}, which shows how 
specific IR line ratios can discriminate between nuclear activity from AGN, normal star formation and also low metallicity star formation. The line ratios differ by orders of magnitude for the different objects, making this diagram very powerful, i.e. it is the analog of the classical optical BPT diagram \citep{1981PASP...93....5B}, 
%(XXX ref.BPT)
but using extinction-free IR lines. 

\begin{figure}[ht]
  \begin{center}
  \includegraphics[width=\columnwidth]{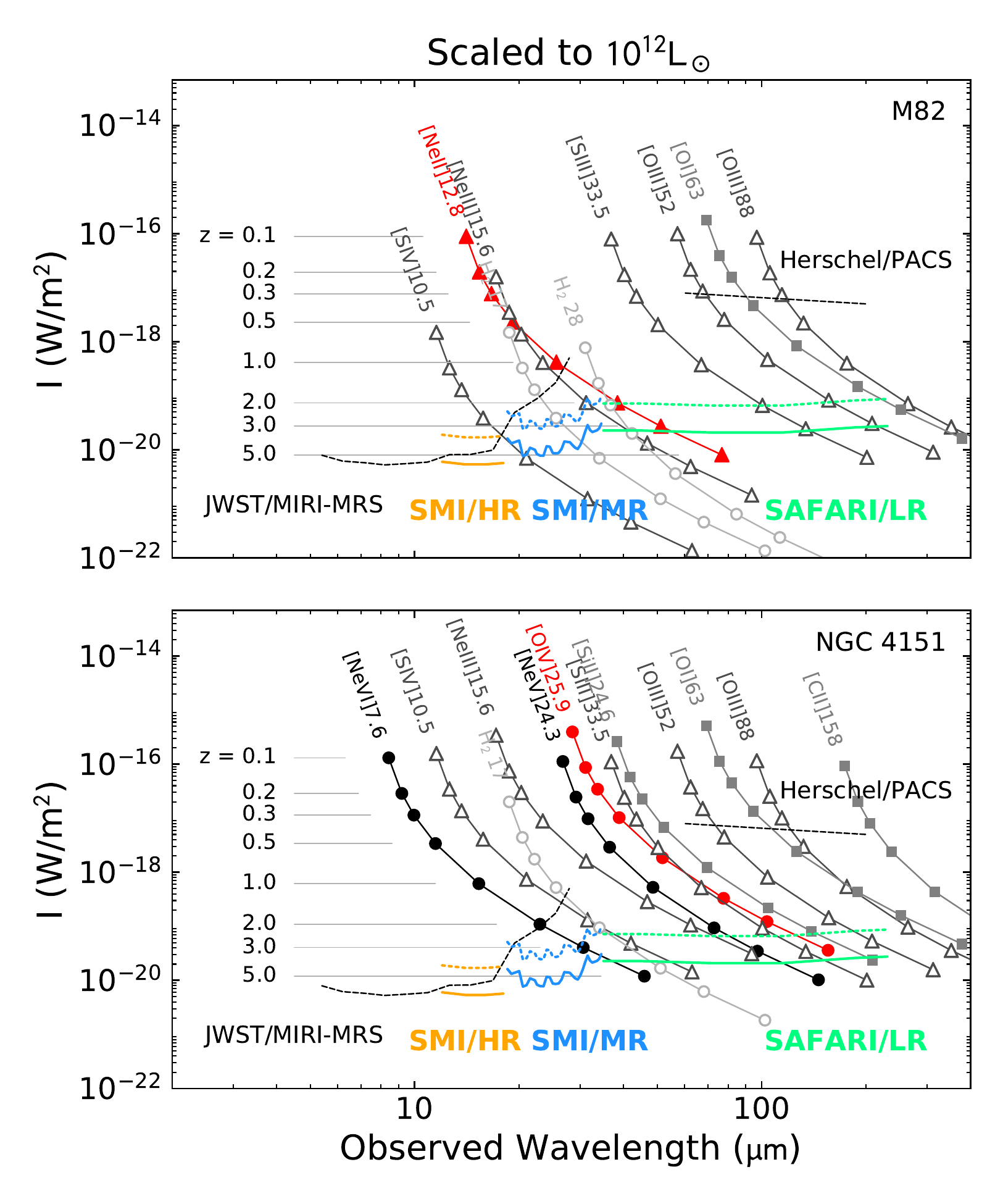}~
\caption{Predicted intensities for the mid- and far-IR fine-structure lines covered by SPICA for the starburst galaxy M82 (upper panel) and the AGN/Seyfert type 1 galaxy NGC\,4151 (lower panel), scaled to a luminosity of $10^{12}\, \rm{L_\odot}$. Filled circles correspond to tracers of AGN activity, triangles to lines typically dominated by star formation, squares to typical PDR lines, and open circles to transitions of warm molecular gas. The predicted line intensities are compared to the $5 \sigma$, 1 hour (dotted lines) and 10 hours (solid lines) sensitivities for SMI/HR (in yellow), SMI/MR (in blue), and SAFARI/LR (in green). Additionally, black-dashed lines indicate the sensitivities for \textit{JWST}/MIRI-MRS ($10 \sigma$, 2.8 hours, from \citealt{gla15}) and \textit{Herschel}/PACS ($5 \sigma$, 1 hour). In less than few hours, \textit{SPICA} will be able to detect the main star-formation and AGN tracers (e.g. [NeII]12.8\,$\mu$m and [OIV]25.9\,$\mu$m) at the peak of star formation and SMBH accretion activity ($1 < z < 3$), for starburst galaxies and AGN at the knee of the luminosity function.} \label{Fig6}
\end{center}
\end{figure}

The SAFARI grating spectrometer at low resolution ({\it R}$\sim$300) will detect [OIV]26\,$\mu$m at $z$=1 in galaxies with {\it L}$\sim$10$^{11}$L$_{\odot}$, at $z$=2 in galaxies with {\it L}$\sim$3$\times$10$^{11}$L$_{\odot}$, and at $z$=3 in galaxies with {\it L}$\sim$10$^{12}$L$_{\odot}$ in a few hours. These luminosities correspond to the knee of the luminosity function, {\it L*} \citep[or characteristic luminosity,][]{1976ApJ...203..297S}, %(Schechter+1976)
at each redshift. Simultaneously, SAFARI will detect many other diagnostic lines of the ionized and neutral ISM, such as the star formation tracers of [NeII]12.8\,$\mu$m, [NeIII]15.5\,$\mu$m, [SIII]18.7\,$\mu$m and 33.4\,$\mu$m, and at longer wavelengths the photodissociation region tracers of [OI]63\,$\mu$m and 145\,$\mu$m and [CII]158\,$\mu$m. 

Fig. \ref{Fig5} shows the IR spectrum of the local active galaxy MCG-3-34-64, rescaled to a luminosity of {\it L}=10$^{12}$L$_{\odot}$ at redshifts $z$ from 1 to 4. This illustrates
that the SPICA spectrometers will be able to detect both the continuum and the brightest lines in such a galaxy up to $z$$\simeq$4 in a few hours. 
Fig.\,\ref{Fig6} shows the predicted intensities for the strongest mid- and far-IR lines in the nearby starburst galaxy M82 and in the AGN NGC\,4151. For all galaxies, the line intensities have been scaled to a luminosity of {\it L}=10$^{12}\, \rm{L_\odot}$. 
The lines of [NeII]12.8\,$\mu$m, [OIII]52, 88\,$\mu$m, %and AGN activity, e.g. 
[OIV]25.9\,$\mu$m, [NeV]14.3\,$\mu$m, 24.3\,$\mu$m and [NeVI]7.6\,$\mu$m %. These lines 
will be detected in less than a few hours for the bulk of both starburst galaxies and AGN (${\it L} \sim 10^{12}\, \rm{L_\odot}$) at the peak of star formation and SMBH accretion activity (1$<$$z$$<$3). 

SPICA will also obtain deep 34--230\,$\mu$m %SAFARI 
low-resolution spectra of around 1,000 galaxies, equally spaced in luminosity and redshift bins, up to a redshift of $z$=3.5 in a total integration time of around 2,000 hrs. Unbiased, deep low-resolution ({\it R}$\sim$50-120) SMI spectrophotometric surveys will be able %are foreseen 
to cover 10 deg$^2$ in 600 hrs. and detect 30,000 galaxies \citep{kan17}.

JWST will not cover wavelengths above $\lambda=28.8{\mu}$m, limiting its capability to measuring spectra of galaxies and AGN above $z$$\sim$2 in the rest-frame mid-IR.  This can be done with JWST %only with the faint lines of, e.g., 
with the [ArII]7\,$\mu$m and [NeVI]7.6\,$\mu$m lines, but these will be a factor of 3--4 fainter than the MIR lines accessible to SPICA.  
As a consequence of this, and incorporating the degradation of JWST MIRI sensitivity toward its long-wavelength band edge, we expect that SPICA will be 25 to 2000 times faster than JWST at diagnosing the power sources in dusty galaxies at $z$=2 and 3, respectively.

%On the other hand, 
ALMA cannot trace far-IR cooling lines (e.g., the [OIII]88\,$\mu$m) at redshifts lower than 3, and cannot probe the dust and gas spectral features 
%visible to SPICA below the observed wavelengths 
shortwards of 300\,$\mu$m, leaving for SPICA the unexplored territory which covers the peak of the SFR and BHAR density functions (1$<$$z$$<$3), at mid- to far-IR wavelengths.

\begin{figure}[h]
\begin{center}
\includegraphics[width=0.9\columnwidth]{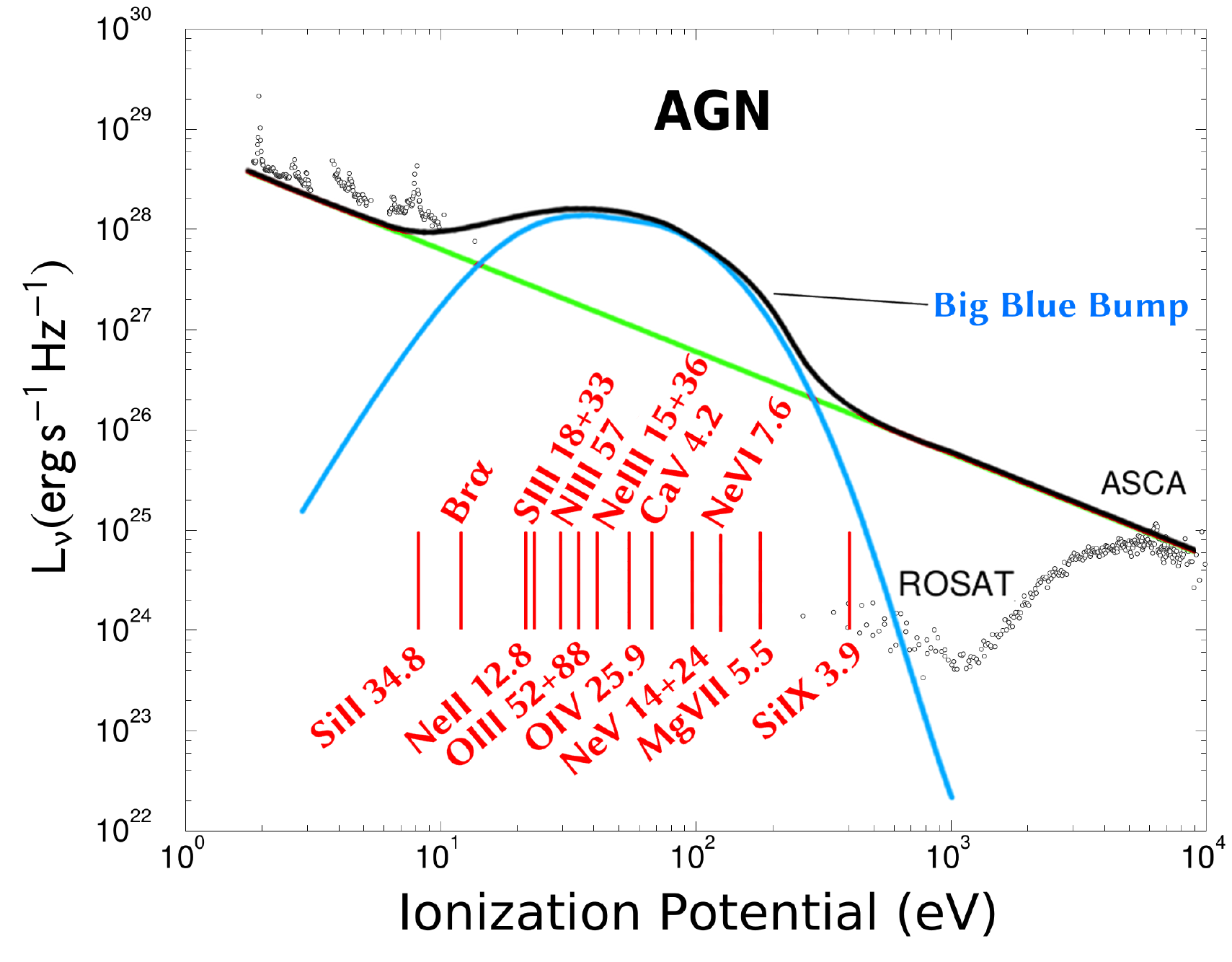}
\includegraphics[width=0.9\columnwidth]{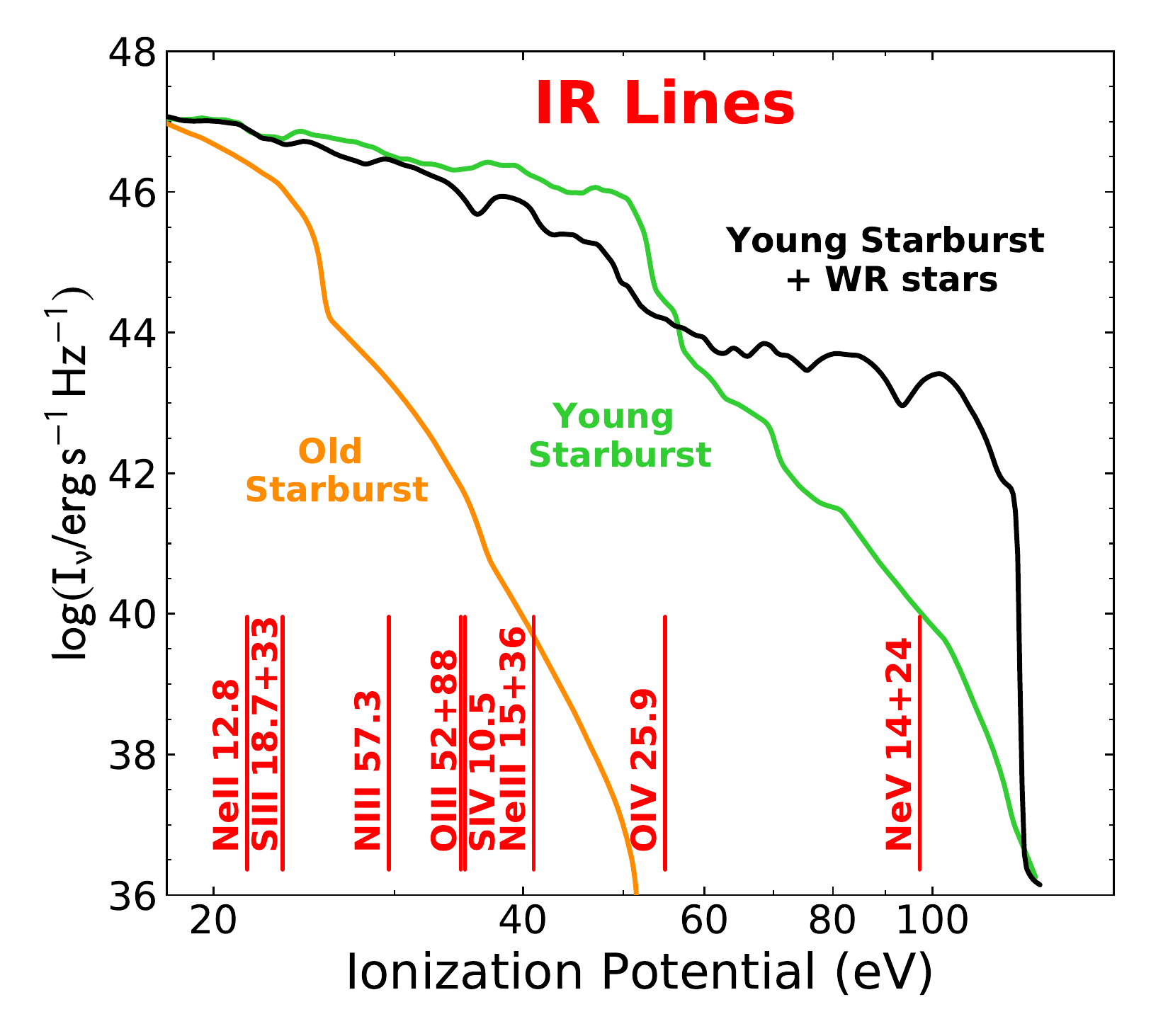}
\caption{Diagrams showing how IR lines, which can be measured by SPICA up to $z$ $\sim$1--2, can map the intrinsic ionizing spectra of both AGN and starburst galaxies.
{\it Upper panel:} Overlay of the NGC4151 primary ionising spectrum (black points) with a schematic "blue bump" and power law 
\citep[adapted from][]{1999ApJ...512..204A}. The IR lines will trace the unobserved primary spectrum at the energies of their ionization potentials. %(adapted from Alexander et al 1999).
 {\it Lower panel:} IR lines probe the starburst's primary ionizing spectrum, which is sensitive to the presence of Wolf-Rayet stellar winds or shocks and to the starburst age \citep[models from][]{1999ApJS..123....3L}. }%(models from Leitherer et al 1999).
 \label{Fig7}
\end{center}
\end{figure}

\subsection{Measuring the primary ionizing spectrum of AGN and starburst galaxies} \label{sect2.1.1}

The intrinsic primary unattenuated ionizing spectrum of AGN or starburst galaxies cannot be observed from the Lyman limit up to several hundred eV, 
because of Galactic and intrinsic absorption. Fortunately, for any region for which the ionization structure is dominated by photons, the primary ionizing spectrum can be inferred from the suite of mid- and far-IR emission lines from the various ions, largely unaffected by dust.
In the nearby Universe, SPICA will detect the full range of mid- to far-IR fine-structure lines, %even the faintest ones, 
providing a powerful probe of the intrinsic, ionizing spectrum. Meanwhile at higher redshift, 
%At intermediate redshifts, 
(e.g. $z$ $\sim$1-2, depending on the luminosity of the galaxy) SPICA will still be able to detect the brightest lines of, e.g., [NeII]12.8\,$\mu$m, [NeIII]15.5\,$\mu$m, [OIII]52 and 88\,$\mu$m, [OIV]26\,$\mu$m and [NeV] 14.3 and 24.3\,$\mu$m, whose ionization potentials are positioned to properly sample the ionizing spectrum.

Fig. \ref{Fig7} (upper panel) shows how the observed IR emission lines trace the primary spectrum of the Seyfert galaxy NGC4151, which is not directly observable from 10 to 200eV,  %\citep{1999ApJ...512..204A}, 
where the signature of the BH accretion disk (big blue bump) should become prominent \citep{1982ApJ...254...22M}. %(Malkan \& Sargent 1982). 
This method has been applied to ISO spectra of bright Seyfert galaxies, e.g. NGC4151 and NGC1068 \citep{1999ApJ...512..204A, 2000ApJ...536..710A}, %(Alexander et al. 1999), 
and {\it Spitzer} spectra of local AGN \citep{2011ApJ...738....6M}. %(Melendez et al. 2011). 
Fig. \ref{Fig7} (lower panel) shows a similar plot for starburst galaxies, where the stellar spectrum can be traced by the IR lines, free from dust extinction. 
This often leads to unexpected results. For example, observations of the [OIV]26\,$\mu$m line have revealed that starburst and dwarf galaxies can have, in particular conditions, a significant contribution of ionizing photons with energies above 50 eV \citep{1998A&A...333L..75L, 1999A&A...345L..17S, 2016ApJS..226...19F}, %(Lutz 1998; Shaerer \& Stasinska 1999; Fernandez-Ontiveros et al 2016), 
which is difficult to reconcile with current stellar population models \citep{2015A&A...576A..83S}. % (Stasinska et al 2015). 
Observations of low-redshift starburst and dwarf galaxies with SPICA will have the potential to reveal the physical mechanisms (e.g. strong winds or shocks, associated 
with very high mass stars) responsible for the high ionization tail of the underlying ionizing spectrum, constraining the upper mass cutoff 
and the stellar initial mass function.

\subsection{Finding AGN in dwarf galaxies} \label{sect2.1.2}

It has been recently recognised that finding and characterising AGN in dwarf galaxies is a unique way to infer the properties of high-redshift BH seeds 
($z$$>$5), i.e. accreting BHs with masses of 10$^5$--10$^6$ M$_{\odot}$, which are probably associated with low-metallicity environments \citep{2016arXiv160907148B}. 
Furthermore, dwarf galaxies are thought to be the local analogues of the high-redshift galaxies responsible for cosmic reionization, 
due to escaping photons produced by strong star formation \citep[Ly  $\alpha$ emitters;][]{2016MNRAS.458L..94S, 2016MNRAS.461.3683I, 2016A&A...591L...8S}
%2014PASA...31...40D}, %(Lymann  $\alpha$ emitters; Izotov+2016, Schaerer+2016, Dijkstra 2014, Sharma et al. 2016), 
and/or by AGN activity from accreting BH of 10$^5$--10$^6$ M$_{\odot}$ \citep{2015ApJ...813L...8M}. %(Madau \& Haardt 2015). 
The IR line ratio diagram shown in Fig. \ref{Fig3}, which will be populated with SPICA spectroscopic observations, is ideal for measuring the AGN contribution 
in dwarf galaxies; this is because dwarf AGN are elusive in X-ray surveys and many of them are contaminated by strong star formation in classical optical $\textquoteleft$$\textquoteleft$BPT"  
\citep{1981PASP...93....5B}
%(XXX ref.BPT)
diagnostics \citep{2016ApJ...829...57B, 2016arXiv160907619S}. %(Baldassare et al. 2016, Simmonds et al. 2016). 
The [NeII] and [NeIII] lines will be detected at low redshift with SMI at high spectral resolution and at increasing redshift with SMI at medium resolution as well as  
with SAFARI.

\section{Synergies with Future Facilities}\label{syner}

SPICA will study the physical processes driving galaxy evolution, %from the near to the distant Universe, %as well as star and planet formation in our Galaxy, 
through sensitive mid- to far-IR observations %, needed in the 
of deeply embedded regions that characterize galaxy formation and evolution at the peak of the SFR and BHAR (1$<$$z$$<$4).
%the birth and most of the evolution of galaxies. %, stars and planets. 
JWST, due to its shorter wavelength range, %will trace warmer material at a more advanced stage and 
will cover the same redshifts at shorter rest-frame wavelengths (e.g. at $z$=3 for $\lambda$$<$7\,$\mu$m and at $z$=2 for $\lambda$$<$ 9\,$\mu$m), %therefore 
missing %, on one side, 
most of the fine-structure diagnostic lines of Table\,\ref{fslines} %. Moreover, due to the increasing dust extinction at shorter wavelengths, it will not trace 
and having a greatly reduced ability to detect most dust-enshrouded and obscured galaxies and AGN, due to the increasing extinction at shorter wavelengths.
%On the overlapping spectral region of SPICA and JWST, between 12-28\,$\mu$m, due to the low temperature and smaller size of its mirror, SPICA will go deeper and will be faster for large fields. % (see Fig. 1-22). 
%However, JWST will be able to 
SPICA will be able to survey large areas of the sky, and find new samples of IR-bright, high-redshift galaxies.  JWST will be able to study
in greater spatial detail the low-redshift Universe ($z$$<$1) and perform cosmological studies of the high-redshift Universe in the UV/optical rest-frame. 
In the submillimeter, ALMA and the NOrthern Extended Millimeter Array (NOEMA\footnote{\url{http://www.iram-institute.org/EN/noema-project.php}}), %NOEMA, 
will map lower excitation emission-line processes and colder dust continuum, with respect to SPICA.
Ground-based interferometry with ALMA and the NOEMA can complement far-IR spectroscopy %SPICA 
in tracing molecular outflows %in ULIRG 
with CO \citep{2010A&A...518L.155F, 2015A&A...583A..99F, 2012A&A...543A..99C, 2014A&A...562A..21C, 2015A&A...580A..35G} %(Feruglio et al. 2010, 2015; Cicone et al. %2012, 2014; García-Burillo et al. (2015)) 
and HCN-HCO$^+$ \citep[e.g.,][]{2009ApJ...700L.104S, 2012A&A...537A..44A, 2016AJ....152..218I}. %(Sakamoto+09, Aalto et al. 2012, 2014, Imanishi+16). 
%covering a complementary earlier stage of evolution of any system, from planets through stars to galaxies. 
%SPICA will thus play a central role in the study of galaxy evolution, %and star and planet formation, 
%because it covers a unique spectral range, inaccessible by any other observatory and tracing the peak of the excitation of many of the most important physical processes. %(which contains a wealth of atomic/ionic as well as molecular transitions and PAH bands)    However, 
SPICA will, however, work in synergy with these facilities and the new generation ELTs, as well as with the X-ray telescope Athena, the \textit{Advanced Telescope for High ENergy Astrophysics} \citep{2013arXiv1306.2307N}. This latter observatory %, foreseen to operate at the same time as SPICA, 
%will be sensitive to the most energetic phenomena in the Universe. It 
will measure, in synergy with SPICA, the BHAR history as a function of cosmic time. However, it will be mostly limited to Compton thin sources, needing SPICA observations to access %derive the complete 
the obscured BHAR density. On AGN feedback, Athena will detect the high energy ionic winds in synergy with SPICA observations of massive molecular outflows.

If approved, SPICA will be operational %would arrive during an exciting period for multi-wavelength astronomy and galaxy evolution studies in particular. This will happen 
in the late 2020s in the era of the ELTs %, just after JWST, with ALMA % Atacama Large Millimeter/submillimeter Array (ALMA; \mbox{\citealt{2009IEEEP..97.1463W}}) and NOEMA
%the NOrthern Extended Millimeter Array (NOEMA\footnote{\url{http://www.iram-institute.org/EN/noema-project.php}}) at their maximum capabilities, 
and at the dawn of the Square Kilometer Array \mbox{\citep[SKA;][]{dew09}}. %Moreover, 
SPICA %will not fly alone, as it will come together with 
will operate at the same time as Athena %the \textit{Advanced Telescope for High ENergy Astrophysics} \citep[Athena;][]{2013arXiv1306.2307N} 
in the X-ray range.
By then, both \textit{Euclid} \citep{mac16} %in the optical and will overlap with the last years of 
and the \textit{Wide Field IR Survey Telescope} \citep[WFIRST;][]{spe13} will have brought a vast legacy of
%in the near-IR range. 
%planned to %be in operate in the early to mid 2020's, 
%Thus, future studies on the %chemical 
%evolution of galaxies will have to take into account the observables and synergies that all these facilities will provide in $\sim 10-15$ years from now.
%In the early to mid 2020's, \textit{Euclid} and WFIRST 
%These will yield a vast legacy of 
deep near-IR images and spectra over large fields, %covering more than half of the sky 
reaching the peak of star formation and SMBH accretion activity  for 1$<$z$<$3, but limited to the unobscured populations of galaxies and AGN. 
From these large surveys, follow-up spectroscopic observations %for a number of candidates %will be performed with JWST and the ELTs and available in the late 2020's. %The lack of dusty and obscured galaxies in these studies could be solved by surveys with both the Athena Wide Field Imager --\,detecting hundreds of mildly Compton-thick obscured AGN up to $z \sim 4$\,-- and the 
will be collected by SPICA mid  and far-IR low-resolution ({\it R} $\sim$ 50--300) spectroscopy,
enabling a full characterization of the physical state of very large samples of galaxies.  

\section{CONCLUSIONS} \label{conc}

The SPICA mission, with its 2.5-m actively-cooled telescope and powerful mid- and far-IR instrumentation, will represent a huge step forward in understanding galaxy formation 
and evolution, through IR spectroscopy of individual galaxies and deep photometric and spectrophotometric surveys
for large statistically significant samples. The mid- and far-IR spectral range is host to a powerful suite of diagnostic tools, able to penetrate even the dustiest regions.  With SPICA astronomers will use this toolset to probe galaxy evolution in several unique ways, including:
\begin{enumerate}
\item[(i)] obtaining the first physical determination through IR spectroscopy of the star-formation rate and of the black hole accretion rate histories across 
cosmic times up to a redshift of $z$$\sim$4;
\item[(ii)] studying AGN accretion and feedback and their impact in the evolution of star formation, through detection and characterisation of far-IR molecular and atomic line profiles;
\item[(iii)] measuring the rise of metals %and dust 
from the distant Universe to the present day through spectroscopy of IR lines, which minimises the effects of dust extinction and eliminates those due to temperature uncertainties and fluctuations in the ionized gas;
\item[(iv)] studying early AGN and starburst-dominated galaxies in significant samples of dusty galaxies up to $z$$\sim$6, through deep (spectro)-photometric surveys; 
\item[(v)] detecting and characterizing some of the youngest and most luminous galaxies in the early Universe when it was only half a billion years old. 
\end{enumerate}

With a large 2000-hr. spectroscopic survey, SPICA/SAFARI will obtain individual high S/N spectra of over 1000 galaxies to $z$$\sim$4. These observations will simultaneously detect diagnostic lines of the ionized and neutral ISM, characterising the local environment and quantifying the contribution from young stars and AGN to the bolometric luminosity. SPICA/SMI will perform similar spectroscopic studies at lower redshifts ($z$$<$2) with low-resolution, wide field surveys, to detect and characterize thousands of dusty galaxies via their PAH features and hot dust emission. No other currently planned telescope will be able to perform this type of detailed spectroscopic and wide-area spectro-photometric investigation. SPICA will fill the wide spectral gap left between ALMA (observing above 300\,$\mu$m) and JWST (below 28\,$\mu$m). In this spectral range, SPICA will observe the ionic fine structure lines, the H$_2$ as well as many other molecular lines (e.g. from H$_2$O, OH, and CO) and PAH features in the rest-frame mid-IR in galaxies at the peak epoch of star formation, as well as those from the molecular gas and photodissociation regions in the far-infrared. 

\begin{acknowledgements}
This paper is dedicated to the memory of Bruce Swinyard, who initiated the SPICA project in Europe, but unfortunately died on 22 May 2015 at the age of 52. He was ISO-LWS calibration scientist, Herschel-SPIRE instrument scientist, first European PI of SPICA and first design lead of SAFARI.
%The SPICA-J Team and 
We acknowledge the SAFARI Consortium and the full SPICA Team, without whose work this project would not have been possible. 
F.J.C. and A.A.-H. acknowledge financial support through grant AYA2015-64346-C2-1-P (MINECO/FEDER).
H.D. acknowledges financial support from the Spanish Ministry of Economy and Competitiveness (MINECO) under the 2014 Ramón y Cajal program MINECO RYC-2014-15686.
T.N. acknowledges financial support by JSPS KAKENHI Grant Number 27247030.
F.N. acknowledges Spanish grants FIS2012-39162-C06-01, ESP2013-47809-C3-1-R and ESP2015-65597-C4-1-R.
Basic research in IR astronomy at NRL is funded by the US ONR.
We thank the anonymous referee, who helped improving the readability of this article. 

\end{acknowledgements}

\begin{appendix}

\section*{Affiliations}
\affil{$^1$ Istituto di Astrofisica e Planetologia Spaziali, INAF, Via Fosso del Cavaliere 100, I-00133 Roma, Italy}
\affil{$^2$ Centro de Astrobiolog\'ia (CSIC-INTA), Dep de Astrof\'isica, ESAC campus, E-28692 Villanueva de la Ca\~nada, Spain}
\affil{$^3$ IPAC, California Institute of Technology, Pasadena, CA 91125, USA}
\affil{$^4$ Sterrenkundig Observatorium, Department of Physics and Astronomy Universiteit Gent, Krijgslaan 281 S9, B-9000 Gent, Belgium}
\affil{$^5$ Department of Physical Sciences, The Open University, MK7 6AA, Milton Keynes, United Kingdom}
\affil{$^6$ Osservatorio Astrofisico di Arcetri, INAF, Largo E. Fermi 5, 50125 Firenze, Italy}
\affil{$^7$ Department of Astronomy and Joint Space Institute, University of Maryland, College Park, MD 20642 USA}
\affil{$^8$ Div. of Physics, Math and Astronomy, California Institute of Technology, Pasadena, CA 91125 \& JPL, Pasadena, CA 91109, USA}
\affil{$^{9}$Laboratoire d'Astrophysique de Bordeaux, Univ. Bordeaux, CNRS, B18N, All\'ee Geoffroy Saint-Hilaire, 33615, Pessac, France}
\affil{$^{10}$Instituto de F\'isica de Cantabria (CSIC-UC), Avenida de los Castros, E-39005 Santander, Spain}
\affil{$^{11}$Laboratoire AIM, CEA/IRFU/Service d'Astrophysique, Universit\'e Paris Diderot, Bat. 709, F-91191 Gif-sur-Yvette, France}
\affil{$^{12}$Blackett Lab, Imperial College, London, Prince Consort Road, London SW7 2AZ, United Kingdom}
\affil{$^{13}$Instituto de Astrof\'isica de Canarias (IAC), C/V\'ia L\'actea s/n, E--38205 La Laguna, Spain}
\affil{$^{14}$Universidad de La Laguna (ULL), Dept. de Astrof\'isica, Avd. Astrof\'isico Fco. S\'anchez s/n, E--38206 La Laguna, Spain}
\affil{$^{15}$Dept. of Earth Science and Astronomy, Grad. Sch. of Arts and Sciences, Tokyo University, 3-8-1 Komaba, Meguro, Tokyo, Japan}
\affil{$^{16}$School of Sciences, European University Cyprus, Diogenes Street, Engomi, 1516, Nicosia, Cyprus}
\affil{$^{17}$Steward Observatory, University of Arizona, 933 North Cherry Avenue, Tucson, AZ 85721, USA}
\affil{$^{18}$Scuola Normale Superiore, Piazza dei Cavalieri 7, I-56126, Pisa, Italy}
\affil{$^{19}$Naval Research Laboratory, Remote Sensing Division, 4555 Overlook Avenue SW, Washington, DC 20375, USA}
\affil{$^{20}$Dipartimento di Astronomia, Universit\'a di Padova, 35122 Padova, Italy}
\affil{$^{21}$CESR, CNRS/Universit\'e de Toulouse, 9 Avenue du Colonel Roche, BP 44346, 31028 Toulouse Cedex 04, France}
\affil{$^{22}$Universidad de Alcal\'a, Departamento de F\'isica y Matem\'aticas, Campus Universitario, E-28871 Alcal\'a de Henares, Madrid, Spain}
\affil{$^{23}$Osservatorio Astronomico di Bologna, INAF, via Ranzani 1, I-40127 Bologna, Italy}
\affil{$^{24}$Institut d'Astrophysique de Paris, 98 bis Bd Arago, Office 117, 75014 Paris, France} 
\affil{$^{25}$European Southern Observatory, Karl Schwarzschild Strasse 2, D-85748 Garching, Germany}
\affil{$^{26}$National Astronomical Observatory of Japan, 2-21-1 Osawa, Mitaka, Tokyo 181-8588, Japan}
\affil{$^{27}$Graduate School of Science, Nagoya University, Furo-cho, Chikusa-ku, Nagoya 464-8602, Japan}
\affil{$^{28}$School of Science, Tokyo Institute of Technology, 2-12-1 Ookayama, Meguro, Tokyo 152-8551, Japan}
\affil{$^{29}$Institute of Space Astronautical Science, Japan Aerospace Exploration Agency, Sagamihara, Kanagawa 252-5210, Japan}
\affil{$^{30}$Institute of Astronomy, The University of Tokyo, 2-21-1 Osawa, Mitaka, Tokyo 181-0015, Japan}
\affil{$^{31}$Astronomy Division, University of California, Los Angeles, CA 90095-1547, USA} 
\affil{$^{32}$Osservatorio Astronomico di Roma, INAF, Via di Frascati 33, I-00040 Monte Porzio Catone, Italy}
\affil{$^{33}$School of Physics and Astronomy, Cardiff University, Queen's Buildings, The Parade, Cardiff CF24 3AA, United Kingdom}
\affil{$^{34}$Dept. of Earth and Space Science, Graduate School of Science, Osaka University, 1-1, Machikaneyamacho, Toyonaka, Osaka, Japan}
\affil{$^{35}$Research Center for Space and Cosmic Evolution, Ehime University, Matsuyama 790-8577, Japan}
\affil{$^{36}$Centro de Astrobiolog\'ia (CSIC/INTA), ctra. de Ajalvir km. 4, 28850 Torrej\'on de Ardoz, Madrid, Spain}
\affil{$^{37}$Department of Astronomy, Graduate School of Science, The University of Tokyo, 113-0033 Tokyo, Japan}
\affil{$^{38}$Kavli Institute for Cosmology, University of Cambridge, Madingley Road, Cambridge CB3 0HA, United Kingdom}
\affil{$^{39}$Cavendish Laboratory, University of Cambridge, 19 J. J. Thomson Ave., Cambridge CB3 0HE, United Kingdom}
\affil{$^{40}$Dipartimento di Fisica e Astronomia, Universit\'a di Bologna, viale Berti Pichat 6/2, I-40127 Bologna, Italy}
\affil{$^{41}$SRON Netherlands Institute for Space Research, Postbus 800, 9700, AV Groningen, The Netherlands}
\affil{$^{42}$Kapteyn Astronomical Institute, University of Groningen, Postbus 800, 9700 AV, Groningen, The Netherlands}
\affil{$^{43}$Observatoire de Gen\'eve, Universit\'e de Gen\'eve, 51 Ch. des Maillettes, 1290, Versoix, Switzerland}
\affil{$^{44}$Dipartimento di Fisica "G. Marconi", Sapienza Universit\'a di Roma, P.le A. Moro 2, 00185 Roma, Italy}
\affil{$^{45}$Department of Physics and Astronomy, University of British Columbia, 6224 Agricultural Road, Vancouver BC V6T 1Z1, Canada}
\affil{$^{46}$Ritter Astrophysical Research Center, University of Toledo, 2825 West Bancroft Street, M. S. 113, Toledo, OH 43606, USA} 
\affil{$^{47}$Max-Planck-Institut f\"ur extraterrestrische Physik, Postfach 1312, D-85741 Garching, Germany}
\affil{$^{48}$Nordita, KTH Royal Institute of Technology and Stockholm University, Roslagstullsbacken 23, SE-106 91 Stockholm, Sweden}
\end{appendix}
% UNCOMMENT THE LINES BELOW IF YOU WISH TO USE BIBTEX
%\bibliographystyle{apj}
%\bibliography{yourbibfile}
%
%\begin{thebibliography}{}
%
%\%bibitem[\protect\citename{Zackrisson et al.}2011]{za11}
%Zackrisson, E., Rydberg, C.-E., Schaerer, D., et al. 2011, ApJ, 740, 13
%
%\end{thebibliography}
%\nocite*{}
  \bibliographystyle{pasa-mnras}
    \bibliography{spica_galevol_draft_paper}
\end{document}